\documentclass[apj]{emulateapj}
\usepackage{apjfonts} 

\usepackage{longtable}
\usepackage{graphicx}

\makeindex
\citeindextrue

\newcommand{\n}{\nodata}

\journalinfo{Astrophysical Journal}
\submitted{Submitted: January 6, 2004; Accepted: March 9, 2004}

\shorttitle{Kinematics of Jets -- III}
\shortauthors{Kellermann et al.}

\begin{document}

\title{Sub-milliarcsecond Imaging of Quasars and Active 
Galactic Nuclei\\
III. Kinematics of Parsec-Scale Radio Jets}


\author{K. I. Kellermann, M. L. Lister\altaffilmark{1,2},
D. C. Homan\altaffilmark{1,3}} 
\affil{National Radio Astronomy Observatory,
520 Edgemont Road, Charlottesville, VA~22903--2475, U.S.A.;
kkellerm@nrao.edu, mlister@physics.purdue.edu, homand@denison.edu}

\author{R. C. Vermeulen}
\affil{ASTRON,
Postbus 2, NL-7990 AA Dwingeloo, Netherlands; rvermeulen@astron.nl}

\author{M. H. Cohen}
\affil{Department of Astronomy, Mail Stop 105-24,
California Institute of Technology,
Pasadena, CA 91125, U.S.A.; mhc@astro.caltech.edu}

\author{E. Ros, M. Kadler}
\affil{Max-Planck-Institut f\"ur Radioastronomie, Auf dem H\"ugel 69,
D-53121 Bonn, Germany; eros@mpifr-bonn.mpg.de,,
mkadler@mpifr-bonn.mpg.de}

\author{J. A. Zensus}
\affil{Max-Planck-Institut f\"ur Radioastronomie, Auf dem H\"ugel 69,
D-53121 Bonn, Germany and National Radio Astronomy Observatory, 520
Edgemont Road, Charlottesville, VA~22903--2475, U.S.A;
azensus@mpifr-bonn.mpg.de}

\author{Y. Y. Kovalev\altaffilmark{1}}
\affil{National Radio Astronomy Observatory,
P.O. Box 2, Green Bank, WV 24944, U.S.A.}
\affil{Astro Space Center of P.N. Lebedev Physical Institute,
Profsoyuznaya 84/32, 117997 Moscow, Russia; ykovalev@nrao.edu}
\altaffiltext{1}{Karl Jansky Fellow}
\altaffiltext{2}{Present Address: Department of Physics, Purdue
University, 525 Northwestern Avenue, West Lafayette, IN 47907, U.S.A.}
\altaffiltext{3}{Present Address:  Department of Physics and
Astronomy, Denison University, Granville,  OH 43023, U.S.A.}


\begin{abstract}

We report the results of a 15~GHz (2~cm) multi-epoch VLBA program,
begun in 1994 to study the outflow in radio jets ejected from quasars
and active galaxies. The observed flow of 208 distinct features
measured in 110 quasars, active galaxies, and BL~Lac objects shows
highly collimated relativistic motion with apparent transverse
velocities typically between zero and about $15c$, with a tail extending
up to about $34c$. Within individual jets, different features appear to
move with a similar characteristic velocity which may represent an
underlying continuous jet flow, but we also see some stationary and
even apparently inward moving features which co-exist with the main
features. Comparison of our observations with published data at other
wavelengths suggests that there is a systematic decrease in apparent
velocity with increasing wavelength, probably because the observations
at different wavelengths sample different parts of the jet structure. 

The observed distribution of linear velocities is not consistent with
any simple ballistic model. Either there is a rather broad range of
Lorentz factors, a significant difference between the velocity of the
bulk relativistic flow and the pattern speed of underlying shocks, or
a combination of these options. Assuming a ballistic flow, comparison
of observed apparent velocities and Doppler factors computed from the
time scale of flux density variations is consistent with a steep power
law distribution of intrinsic Lorentz factors, an isotropic
distribution of orientations of the parent population, and intrinsic
brightness temperatures about an order of magnitude below the
canonical inverse Compton limit. It appears that the parent population
of radio jets is not dominated by highly relativistic flows, and
contrary to the assumption of simple unified models, not all sources
have intrinsic speeds close to $c$.

Usually, the observed jet flow is in the general direction of an
established jet. However, many jets show significant bends and twists,
where the observed motions are non-radial, but are alingned with the local
jet direction suggesting that the jet flow occurs along pre-existing
bent channels.  In a few cases we have observed a clear change in the
direction of a feature as it flows along the jet.  Radio jets which
are also strong gamma-ray sources detected by EGRET appear to
have significantly faster speeds than the non EGRET sources,
consistent with the idea that gamma ray sources have larger Doppler
factors than non gamma-ray sources. Sources at high redshift have
systematically lower angular speeds than low redshift jets, consistent
with standard cosmologies.

\end{abstract}


%

\section{INTRODUCTION}
\label{intro}

The discovery in the mid 1960s of rapid variability in extragalactic
radio sources \citep{S65,D65,PK66} appeared difficult to explain in
terms of conventional synchrotron radiation theory. These early
observations showed changes by as much as 25 percent over a few weeks
as well as significant day-to-day variations \citep{PK66}. It was
quickly realized (e.g., \citealt{HBS66}), that the observed rapid
variability implied such small linear dimensions, that the relativistic
electron population would be rapidly extinguished by inverse Compton
scattering. Later, \cite{KP69} put these arguments on a quantitative
observational basis showing that due to inverse Compton cooling, the
maximum sustainable brightness temperature for an incoherent source of
electron synchrotron radiation is less than $10^{12}$~K.

However, \cite{W66} and \cite{R66,R67} pointed out that if there is
relativistic bulk motion of the emitting material, since the radiation
is then beamed along the direction of motion, the apparent luminosity
in that direction is enhanced, while at the same time the cross-section
for inverse Compton scattering is reduced. Additionally, since the
radiating source nearly catches up with its radiation, for a favorably
positioned observer, the apparent transverse motion can exceed the
speed of light. Early VLBI observations \citep{W71, CCP71} showed
evidence for the predicted high-velocity outflow; however, the
arguments were indirect, and one had to have faith in the
interpretation of the very limited radio interferometric data which did
not fully sample the source structure (e.g., \citealt{D72}). Subsequent
higher quality VLBI observations confirmed the existence of
superluminal motion in the well-collimated radio jets, found in the
nuclei of many quasars as well as in nearby active galaxies
\citep{CLM77}.

VLBI observations provide a direct method to investigate aspects of the
formation, acceleration, and propagation of extragalactic radio jets.
Early studies discussing apparent jet speeds were statistically
unreliable because they were largely based on the analysis of published
observations of only a few tens of sources made at different times,
often at only two or three epochs, by different groups, using different
antenna array configurations, and different observing/data reduction
strategies. The dynamic range of the images was often inadequate to
identify individual features, especially for the more complex sources,
and spacings in time between successive observations were frequently
too long to uniquely identify and track moving jet features from epoch
to epoch. Moreover, the source selection criteria for many previously
published studies of superluminal motion were not well defined. A
systematic study by \cite{V95} of 81 sources from the flux density
limited 6~cm Caltech-Jodrell Bank Survey indicated smaller typical
speeds than in earlier reports.  This suggested that earlier studies
preferentially tended to observe and/or report only sources in which
rapid motion was detected or suspected.

In 1994 we began a systematic 15~GHz VLBA survey of relativistic
outflows in a sample of over one hundred quasar and active galaxy radio
jets. Our motivation was to study the distributions of velocities,
bending, pattern motions, accelerations, and other complexities of the
jet kinematics that may exist, as well as changes in the strength and
morphology of features as they propagate along the jet. These
observations have provided the homogeneity, resolution and dynamic
range to reliably distinguish and identify individual components
between different epochs. The central VLBA antennas provide short
interferometric spacings that have allowed us to a) track features
moving for some distance down the jet where they become more diffuse,
and b) observe the continuous jet rather than just the bright features
which are often referred to as ``components'' or ``blobs''. 

We chose 15~GHz as a compromise between achieving the best angular
resolution, and the better sensitivity and immunity from weather
conditions found at lower frequencies. At this observing frequency the
resolution of the VLBA is approximately 0.5~mas in right ascension and
between 0.5 and 1~mas in declination. We therefore have sufficient
angular resolution in many cases to resolve the 2-dimensional jet
transverse to its flow, which allows for tests of some theoretical
models. Also, individual features can often be recognized at
significantly smaller separations from the origin. On the other hand,
our observations are less sensitive to structure and motions of the
more diffuse features located far out along the jet or within 0.5~mas
of the core.

Our 15~GHz data represent a significant improvement over previous
studies of superluminal motion in AGN. First, our sample of sources is
large, and the sample membership was based on criteria other than
observed superluminal motion. Second, the speeds are better determined
because a) our data are well-sampled and span a longer time period, and
b) our higher image resolution results in less blending of features.

In Paper I of this series \citep{K98A}, we described the parsec-scale
structure of 132 of the strongest known radio jets based on a single
epoch of observation. In Paper II \citep{Z02} we discussed the
structure of an additional 39 sources. Contour maps of all of our
multi-epoch observations are available on our web site
\footnote{http://www.nrao.edu/2cmsurvey/} along with kinematic and
other data on each source. In this paper we discuss our multi-epoch
observations made through 15~March 2001 and the kinematics derived from
these observations.

Some preliminary results of our program have already been published
(\citealt{KVZ99, K00, K02, RPL02, CRH03, Kel03, K03, Z03}). We have also
discussed the kinematics of several individual sources including
PKS\,1345+125 \citep{LKV03}, NGC\,1052 \citep{VRK03} and 3C\,279
\citep{HLK03}.

In \S~\ref{sampledef} we describe our sample and source selection
criteria, and in \S~\ref{observations} we discuss the details of our
observational program. In \S~\ref{beaming} we summarize the predictions
of relativistic beaming models while in \S~\ref{kinematics} we present
the observed jet kinematics. In \S~\ref{beamingmodels} and
\S~\ref{comparison} we discuss the implications of our observations for
the nature of the relativistic flow. In \S~\ref{muz}, we discuss the
angular velocity -- redshift relation, and in \S~\ref{summary} we
summarize our results and describe our on-going observing program.

Throughout this paper we use the following cosmological parameters:
$H_0 = 70$~km\,s$^{-1}$\,Mpc$^{-1}$,
$\Omega_\mathrm{m}=0.3$, and $\Omega_\Lambda=0.7$.

\section{SAMPLE DEFINITION}
\label{sampledef}
\subsection{\it The Full Sample \label{fullsample}}

Our original source sample (see Paper I) was based on the K\"uhr 1~Jy
catalog \citep{KWP81} as supplemented by \cite{SMK94}. Our goal was to
include all known sources in the Stickel et al.\ catalog that have a
flat spectral index ($\alpha>-0.5$ for $S_\nu\sim\nu^\alpha$) anywhere
above 500~MHz, and a total flux density at 15~GHz, observed at least at
one epoch, greater than 1.5~Jy for sources north of the celestial
equator, or greater than 2~Jy for sources between declinations
$-20\arcdeg$ and the equator. Since the Stickel et al.\ catalog is
complete only at 5~GHz, we used other measurements at 15~GHz or
extrapolations from lower frequencies to augment the K\"uhr sample.

While we attempted to be complete for flat-spectrum, core-dominated
sources which met our selection criteria, we also included a number of
other sources of special interest as follows:

a) Six compact steep spectrum (CSS) sources as indicated in
Table~\ref{sample}. These sources generally have steep radio spectra
and angular sizes smaller than $\sim 1\arcsec$ on the sky. 

b) Four lobe-dominated sources with core components that would satisfy
our criteria if they did not also have strong extended steep spectrum
structure. Two other sources, NGC\,1052 (0238$-$084) and 3C\,120
(0430+052), also have prominent double lobe structure, but since they
are core-dominated on arcsecond scales at 15~GHz, they met our spectral
index selection criterion.

c) Ten Gigahertz-peaked spectrum (GPS) sources, some of which have
two-sided jet structure (\citealt{LKP02}; M.~L.\ Lister et al., in
preparation). However, many of the GPS sources we observed turned out
to be merely flat-spectrum sources whose spectrum was temporarily
dominated by a bright, synchrotron self-absorbed component in the jet.
Our ``peaked'' classification in Table~\ref{sample} is given only to
sources that to our knowledge have always met the GPS spectral criteria
given by \cite{DBO97}. For a few sources, our classification differs
from that previously published (e.g., \citealt{OD98}); however we
believe that our classification is more robust (\citealt{K04} and
Y.~Y.\ Kovalev et al., in preparation).

Table~\ref{sample} summarizes the properties of each source discussed
in this paper. All of these sources were introduced in Papers I and II;
however, 37 sources included in Papers I and II are not discussed now
because they were observed only once or twice. These have subsequently
been observed and will be discussed in a separate paper (E.\ Ros et al., in
preparation). One source listed in Paper~I is gravitationally lensed
(0218+357), so we have chosen not to include it in our statistical
discussion. Eight other sources had no detectable jet above our
sensitivity limit. They are labeled as ``naked cores'' in
Table~\ref{sample} and we report no motions for these sources. Three
sources, 0316+162, 1328+254, and 1328+307 reported in Papers I or II,
are not discussed here as they have complex structure that we could not
adequately image.

Columns 1 and 2 of Table~\ref{sample} give the IAU source designation
and where appropriate a commonly used alias. Column 3 indicates whether
or not the source is a member of the representative flux density
limited sample that we describe in \S~\ref{subsample}. The optical
classification and redshift as given mainly by \cite{VCV01} are given
in columns 4 and 5, respectively. In column 6 we give a radio spectral
classification for each source based on \mbox{RATAN--600} telescope
observations of broad-band instantaneous spectra from 1 to 22~GHz as
described by \cite{KNK99}. These spectra are available on our web site.
For a few sources not observed at the \mbox{RATAN--600} telescope, we
used published (non-simultaneous) radio flux densities taken from the
literature. We consider a radio spectrum to be ``flat'' if any portion
of its spectrum in the range 0.6 to 22~GHz has a spectral index flatter
than $-0.5$ and steep if the radio spectral index is steeper than $-0.5$
over this entire region. Column 7 shows the parsec scale radio
morphology taken from Papers I and II; in column 8 the largest
total flux density seen on any of our VLBA images, and in column 9 an
indication of whether or not the radio source is associated with a
gamma-ray detection by EGRET. 

\subsection{\it The Representative Flux Density Limited Subsample \label{subsample}}

The full sample of sources described in the previous section is useful
for investigating jet kinematics in a cross-section of known AGN
classes. However, in order to compare with the theoretical predictions
of relativistic beaming models, a well-defined sample selected on the
basis of beamed (not total) flux density is needed. Past surveys (e.g.,
\citealt{TVR96,L01}) have attempted to accomplish this by means of a
spectral flatness criterion. However, we found that this method
eliminates some lobe-dominated active galaxies such as those described
in \S~\ref{fullsample}. Also, we found that the extrapolated 15~GHz
flux density based on non-simultaneous lower frequency measurements was
often grossly in error, because of spectral curvature or variability.

We have therefore assembled a flux density-limited subsample from
the full 15~GHz VLBA survey by using our measured VLBA flux densities as
the main selection criterion. All sources that had a total CLEAN VLBA
flux density exceeding 1.5~Jy (2~Jy for southern sources) {\it at any
epoch since 1994} are included in this sub-sample. We excluded any
sources that were observed on at least four occasions and never
exceeded this limit. For survey sources with fewer than four VLBA
epochs, we estimated the VLBA flux density at various epochs during
this period using the source compactness and data from the flux density
monitoring programs at the \mbox{RATAN--600} radio telescope
\citep{KNK99} or the University of Michigan Radio Astronomy
Observatory\footnote{http://www.astro.lsa.umich.edu/obs/radiotel/umrao.html}
(\citealt{AAH92,AAH03}).

It is important to note that this sub-sample, although flux
density-limited, is not complete; that is, there are additional compact
sources that fulfill our selection criteria. This is partly due to the
lack of a complete all-sky survey at 15~GHz, and also to the variable
nature of AGN. We have therefore identified a list of candidate sources
from other recent high-frequency radio surveys, made since our original
list was compiled in 1994. These include the WMAP survey \citep{Ben03},
the VLBA calibrator survey \citep{BGP02}, \mbox{RATAN--600} observations
\citep{KNK99}, and the high-frequency peaker survey \citep{DSC00}. We
have made new VLBA observations of these sources in order to assemble
a complete flux-density limited, core-selected sample. There are 133
sources in this complete flux density limited sample, but so far we
have multiple epoch observations and derived speeds for only 71 of
these sources which we define as the representative sub-sample. For
the purposes of the statistical analysis presented here, we consider
our present sub-sample to be {\it representative} of a complete
sample, since the general properties of the missing sources should not
be substantially different from the whole. We have compared the 15~GHz
luminosity distributions of our representative sub-sample and the
missing sources from the full sample using a Kolmogorov--Smirnov
(K--S) test, and find no significant difference between the two
samples.

Although our sub-sample selection method is somewhat complex, it is
based on the directly-measured compact flux density, and does not use
an often unreliable spectral index criterion. Also, since the survey
membership is not determined from a single ``snapshot'' epoch, we are
not excluding potentially interesting or highly variable sources simply
because they happened to be in a low state at the time of the original
investigation. This increases the size of the sample and the robustness
of statistical tests on source properties. Of the 71 sources in our
representative sub-sample, there are 53 quasars, 12 BL~Lac objects, and
6 galaxies.

\section{THE OBSERVATIONS AND ANALYSIS}
\label{observations}

Our observations were made during 29 separate observing sessions
between 1994 and 2001. Typically, we obtained images of each source at
three to seven epochs over this seven year period.  Sources were
usually observed at 6 to 18 month intervals.  Those with known rapid
motion were observed more frequently, while those with no observed jet
or with only small observed motions were observed less often.  Each
observing session lasted between 8 and 24 hours.  In general we tried
to observe only at elevations above $10\arcdeg$ to minimize the effect
of tropospheric absorption, phase errors, and excessive ground
radiation. Except for sources at low declinations, where the hour
angle coverage is restricted, we observed each source once per hour
for six to eight minutes over a range of eight hours in hour angle.
In general, all ten VLBA antennas were used for each observations,
except when restricted by elevation.  We rarely used less than 8
antennas.  Papers I and II, as well as our web site, give logs of the
observations.

Some additional observations were made in 1998 and 1999 by L.~I.\
Gurvits et al.~(in preparation) as part of a different program to
compare 15~GHz source structure measured with the VLBA to the 5~GHz
structure measured using the Japanese {\it HALCA} space VLBI satellite
\citep{H98}. We have used these data to supplement our own as the
observations were made using the same observing and data reduction
procedures as used for the present program. These images are also
available on our web site.

Data reduction was done using a combination of AIPS and DIFMAP as
described in Paper I\relax. Each image was analyzed using the AIPS
tasks JMFIT or MAXFIT to determine the relative positions of each
definable feature at each epoch, and these positions were then used to
calculate velocities relative to a presumed core component. We have
assumed that the bright unresolved feature typically found at the end
of so-called core-jet sources is the stationary core. Generally, the
choice of the core is clear due to its high brightness temperature and
location. However, in a few cases, the location of the core is
ambiguous, particularly in some sources where the component motions
appear two-sided about a centrally-located core. As an aid in
identifying components from one epoch to the next we examined each
image for continuity in position, flux density, and structure. Usually,
this procedure is more reliable than fitting independent models to the
($u$,$v$) data at each epoch, and with the possible exception of a few
isolated cases, we believe that we have correctly cross-identified
features as they evolve from epoch to epoch. However, for those sources
where the jet is barely resolved from the VLBA core, we fit models to
the data in the ($u$,$v$) plane in order to exploit the full
interferometric resolution of the VLBA. 


\begin{figure*}[p]
\figurenum{1a}
\begin{center}
\resizebox{1.0\hsize}{!}{\includegraphics[trim=0cm 0cm 0cm 0cm,clip]{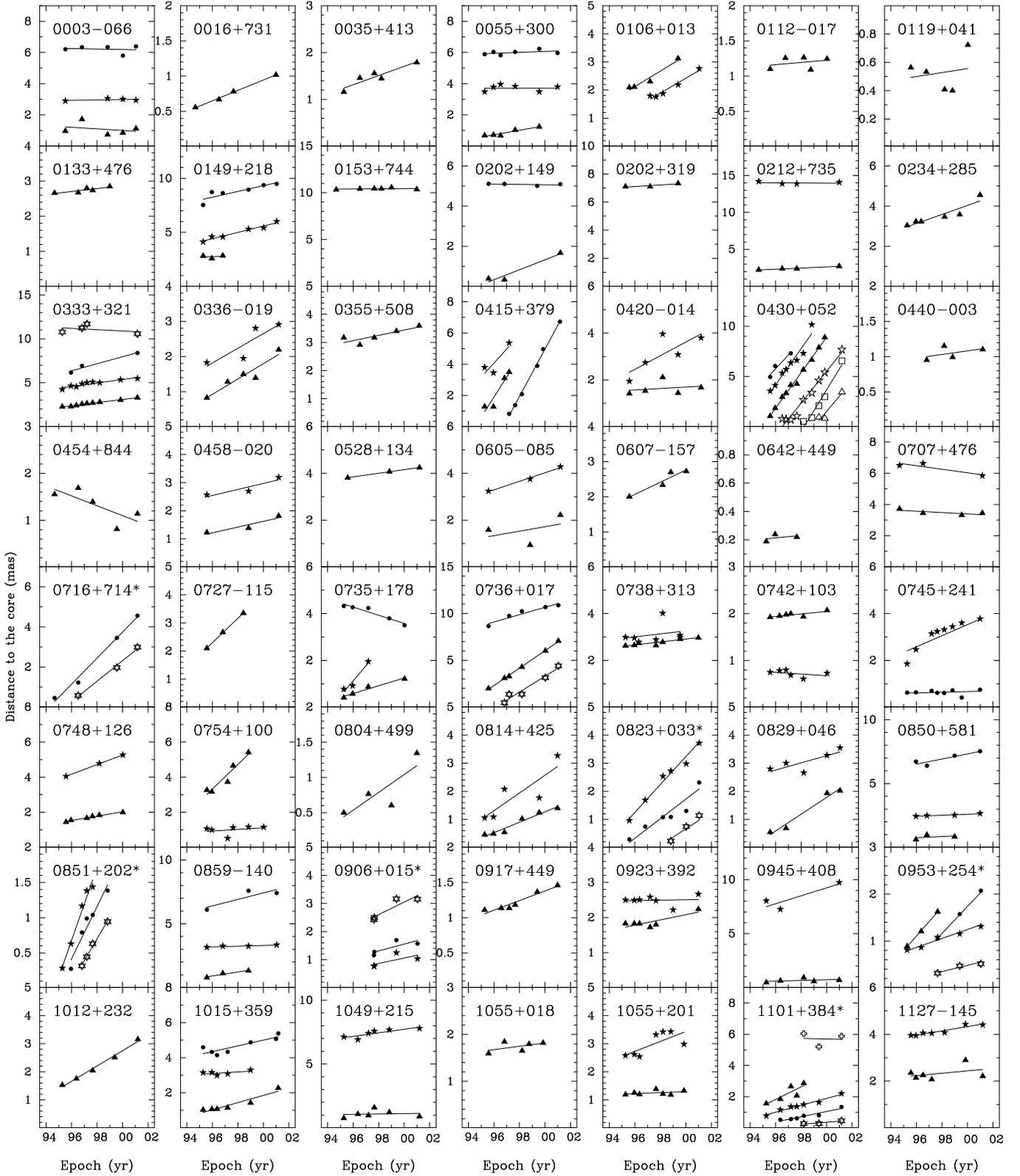}}
\end{center}
\caption{\footnotesize \label{rvt1}
Plots showing the change in separation with time of features of
one-sided jets for which we have measured a velocity from observations
at three or more epochs. An
asterisk denotes sources that were model fit in the ($u$,$v$) plane
rather than in the image plane. Different symbols are used for each
component as follows: B, filled triangle; C, filled five point star; D,
filled circle; E, filled six point star; F, cross; G, five point star;
H, square; I, triangle; J, circle with cross; K, circle with dot; L,
circle. The solid lines denote the best least square linear fit to the data,
and the slope represents the proper motion, $\mu$, tabulated in
Table~\ref{motions}.
}
\end{figure*}

\begin{figure*}[p]
\figurenum{1b}
\begin{center}
\resizebox{1.0\hsize}{!}{\includegraphics[trim=0cm 0cm 0cm 0cm,clip]{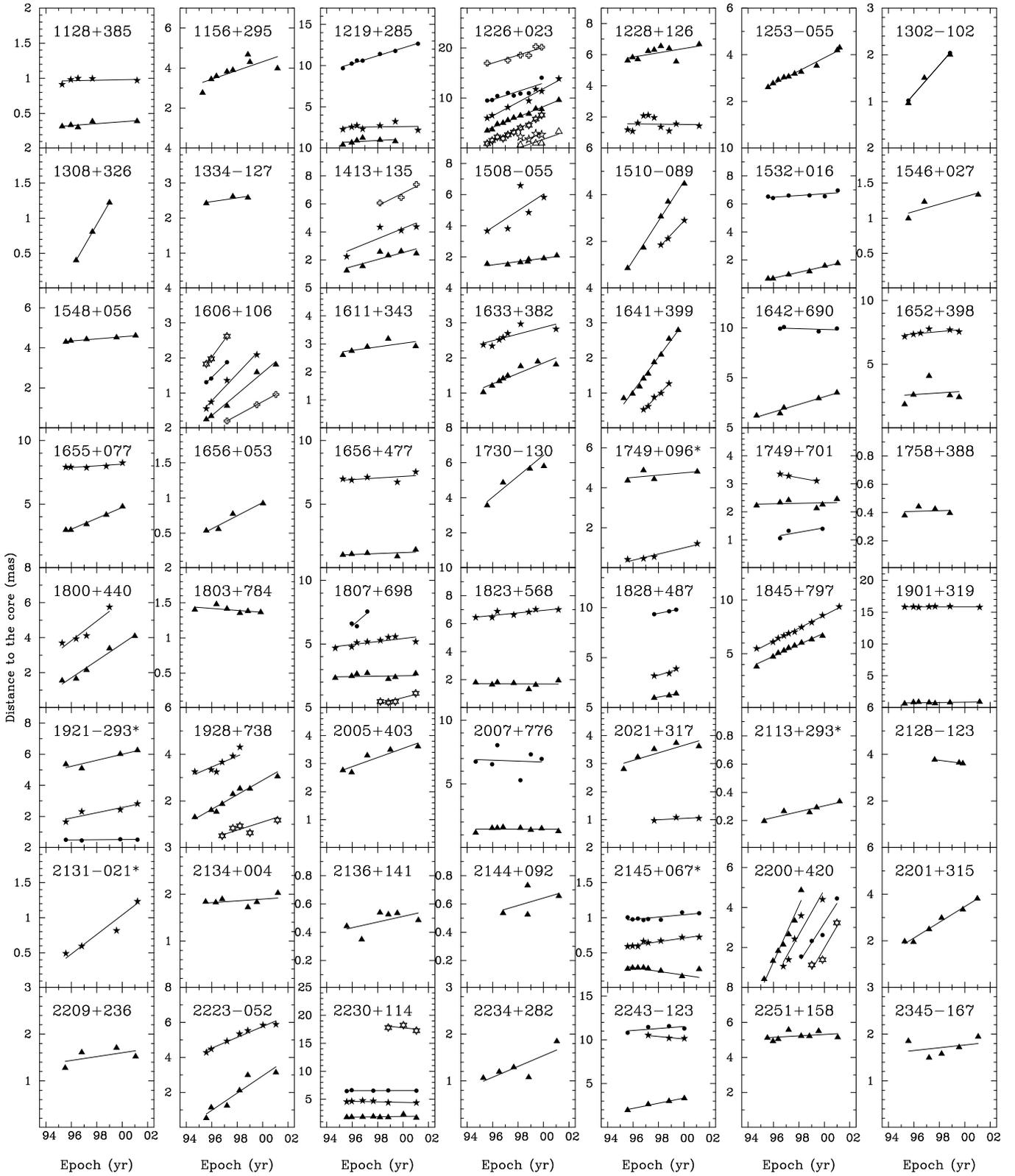}}
\end{center}
\caption{Same as Figure 1a.}
\end{figure*}

\setcounter{figure}{1}
\begin{figure}[t]
\begin{center}
\resizebox{1.00\hsize}{!}{\includegraphics[trim=0cm 0cm 0cm -0.12cm,clip]{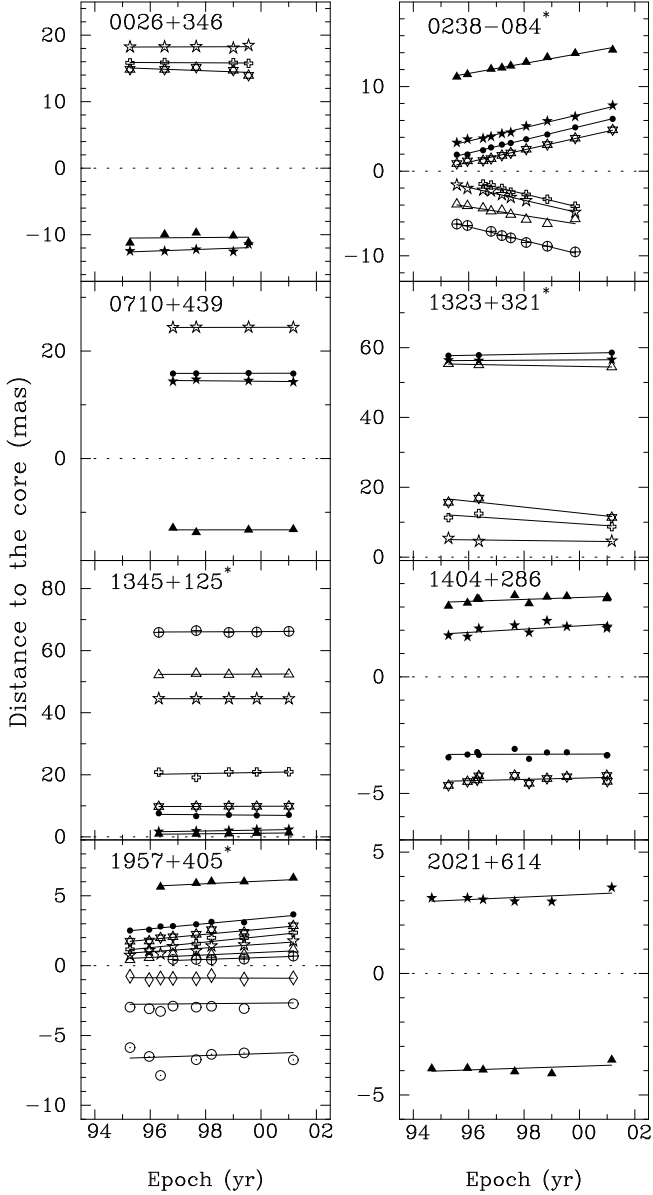}}
\end{center}
\caption{\label{cso}
Same as Figure~1 but for two sided sources only. The points shown for
0238$-$084 (NGC 1052) are the brightest four components located on
each side of the center of symmetry. No central component is visible
for these sources, so the measured component positions are referred to a
virtual center of symmetry as discussed by \cite{VRK03}.}  

\end{figure}

In Figures~1 and~\ref{cso} we show the angular motion in the jets of
the 120 sources for which we have been able to determine motions with
three or more epochs of observation. Of these 120 sources, there are
110 sources with good quality data and measured redshifts for which we
have been able to determine the linear velocity of at least one jet
feature. These comprise 13 active galaxies, 79 quasars, and 18 BL~Lac
Objects. In all, we have been able to determine reliable values for
208 separate features found in these 110 jets.

     We have determined the radial angular speed, $\mu$, of each
definable jet feature using a linear least squares fit to the measured
component positions, relative to the presumed core. Our measured
values of $\mu$ are given in Table~\ref{motions}. In Figures~1
and~\ref{cso} the slope of the line corresponds to the best fitting
least squares fit to the speed which is shown in column 6 of
Table~\ref{motions}. Figure~1 shows the plots for the one-sided jets
and Figure~\ref{cso} the two-sided ones. For well-defined components
the formal uncertainty in the relative position is small, generally
less than 0.02~milliarcseconds (mas). Frequently, however, the jets
have a complex brightness distribution with regions of enhanced
intensity that can brighten and fade with time. In some sources,
instead of a well-collimated jet, there is a broad plume or more
complex two dimensional structure. Not only is the centroid of these
components poorly defined, but changes in the brightness distribution
with time complicate the definition of their motion. Some features
appear to move; while others appear stationary. Some may break up into
two or more separate features, and it is often unclear how these
moving features are related to the underlying relativistic flow. In
some sources, the moving features are not well resolved from each
other, especially when they are close to the core. These issues all
affect our ability to make a unique identification of components from
epoch to epoch. In a few cases where there are fast moving components,
our observations may be too widely spaced in time to uniquely define
their speed. Ambiguities exist, but we believe that most of our
component identifications are robust and that component speeds are
reliably determined.

We have based our error estimates of the angular velocity on the
dispersion about the best-fit linear relation between the measured
feature positions and time. The accuracy of each fit depends on the
strength and size of the feature, the number of epochs, and the range
of time covered by the observations. We note, however, that the
measured dispersion of points about the best-fit line will overestimate
the true scatter in component positions if there is acceleration such as
observed for 3C\,279 \citep{HLK03}. Additionally, the measured radial
speed may be an underestimate of the true speed if there is a
significant non-radial component to the motion. However, we show in
\S~\ref{non-radial-bent} that while non-radial motions are common, the
vector speeds do not differ significantly from the radial speeds,
except in the case of 1548+056. 

We use the following criteria to classify the quality of each measured
velocity.

\begin{itemize}
\item The component position is determined at four or more observations.
\item The component is a well-defined feature whose position can be
unambiguously determined to a small fraction of the VLBA beamwidth.
\item The best fitting angular speed is determined to high accuracy,
defined by an uncertainty $\leq 0.08$~mas per year or a
significance\footnote{The $\geq 5\sigma$ requirement is necessary to
accommodate nearby sources with very high angular speed features which
may have large absolute uncertainties.} $\geq 5\sigma$.
\end{itemize}
We then assign a ``quality code'' to each component motion as follows.

\begin{itemize}
\item E (Excellent) denotes motions that satisfy all three of the above
criteria.
               
\item G (Good) denotes motions that satisfy any two of the above criteria.

\item F (Fair) denotes motions that satisfy only one of the above criteria.

\item P (Poor) denotes motions that do not satisfy any of the above criteria,
or that the uncertainty in the fitted speed is $>0.15$~mas per year
(except for the $\geq 5\sigma$ cases described above).

\end{itemize}

These codes and the measured proper motions are tabulated in
Table~\ref{motions}.

\section{DISCUSSION\label{discussion}}
\subsection{\it Superluminal Motion and Relativistic Beaming \label{beaming}}

We interpret the structural changes seen in the AGN of our sample in
terms of the twin relativistic jet model of \cite{BR74}. In this
framework, the bulk velocity of the relativistic flow, in units of the
speed of light, $c$, is usually denoted as $\beta_\mathrm{b}$, and
$\gamma_\mathrm{b}$ is the corresponding Lorentz factor defined by
$\gamma_\mathrm{b} = (1 - \beta_\mathrm{b}^{2})^{-1/2}$. For a jet flow
at an angle $\theta$ to the observer's line of sight, the Doppler
factor, $\delta$, is given by $\delta= \gamma_\mathrm{b}^{-1}(1 -
\beta_\mathrm{b}\cos{\theta})^{-1}$.

Three important observational consequences of relativistic
motion in synchrotron sources are:

a) A Doppler frequency shift of the radiation, where the ratio of
observed frequency, $\nu$, to the emitted frequency $\nu_{e}$ is
given by
\begin{equation}
  \nu/\nu_\mathrm{e} = \delta.
\end{equation}

b) A change in the observed transverse velocity due to the apparent
time compression. The apparent transverse velocity, in units of $c$, is
given by
\begin{equation}
 \beta_\mathrm{app} = \beta_\mathrm{p} \sin{\theta} /(1 -
 \beta_\mathrm{p} \cos{\theta}),
\end{equation}
where $\beta_\mathrm{p}= v_\mathrm{p}/c$ is the pattern
velocity. We make this distinction between pattern and bulk velocities
because jet features can move with significantly different velocity
than the bulk jet flow (see, e.g., \citealt{LB85, ZCU95, VC94}). This could
be the case, for example, if the pattern motion is due to the
propagation of shocks. By studying 25 core-dominated quasars selected
from the literature, \cite{VC94} showed that the simplest model, which
has only one value of $\gamma$ that is the same in all sources and also
has $\beta_\mathrm{b}=\beta_\mathrm{p}$ in all sources, is not
tenable. They showed that either there must be a distribution of
$\gamma$ among the sources; or that $\beta_\mathrm{b}
\neq \beta_\mathrm{p}$.  Most likely, there is both some distribution
of $\gamma$, and also a difference between $\gamma_\mathrm{p}$ and
$\gamma_\mathrm{b}$

The apparent transverse velocity of an approaching component with a
pattern Lorentz factor $\gamma_\mathrm{p}$, reaches a maximum apparent
speed $\beta_\mathrm{p}\gamma_\mathrm{p}$ when the component moves at
an angle $\sin(\theta) = 1/\gamma_\mathrm{p}$ to the line of sight.
For the corresponding component in the receding jet, the
observed velocity is $\beta_\mathrm{p}/2$.

c) A change in the apparent flux density, $S$, of a moving component
over its stationary value, $S_0$, by a factor
\begin{equation}
S/S_0 = \delta^{x-\alpha},
\label{eq_ss0}
\end{equation}
where $\alpha$ is the spectral index and $x$ has a value of 2 for a
continuous jet or 3 for discrete components (see e.g., \citealt{UP95}). 

The Doppler factor is sharply peaked along the direction of motion, so
sources with highly relativistic jets that happen to be pointed close
to the line of sight will appear strongly boosted, and hence likely to
be selected in a flux-limited sample. Although the jets may be
intrinsically two-sided, unless they are very close to the plane of the
sky, they will appear highly asymmetric (i.e., one-sided) since the
radiation from the receding jet is highly beamed away from the
observer. 

Equation (\ref{eq_ss0}) has the important consequence that the
strongest, most compact radio sources we observe tend to have highly
relativistic jets that are aligned close to the line of sight, and they
will therefore likely display superluminal motion. The most probable
angle for sources selected on the basis of beamed flux density alone is
close to $(2\gamma)^{-1}$, where $\beta_\mathrm{app} \sim \gamma/2$
\citep{VC94}. The distribution of observed speeds and flux densities
is dependent on the distributions of $\gamma$, redshift, and the
intrinsic luminosity function.

\begin{figure}[t]
\begin{center}
\resizebox{1.0\hsize}{!}{\includegraphics[trim=0cm 0cm 0cm -0.2cm,clip]{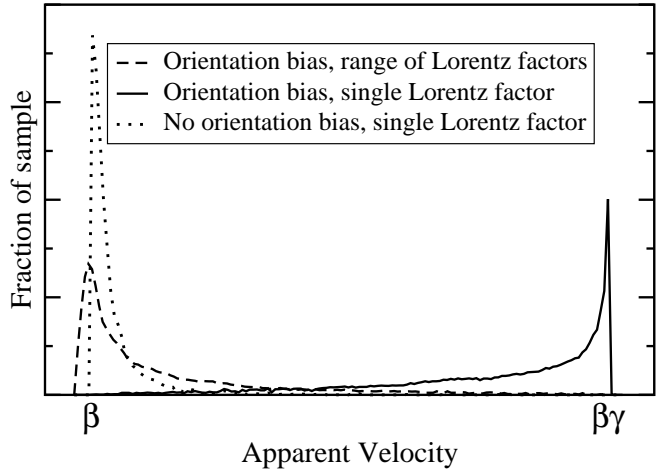}}
\end{center}
\caption{\label{rene}
Predicted distributions of $\beta_\mathrm{app}$ when the sample has
randomly oriented jets all with the same Lorentz factor $\gamma$
(dotted line), when the sample has sources drawn from such a jet
population, but based on flux density which can be enhanced by Doppler
boosting (solid line), and for a Doppler boosting with a range of
Lorentz factors which favors low values of the Lorentz factor (dashed
line).  }
\end{figure}

This is illustrated in Figure~\ref{rene} where we show the predicted
distribution of apparent velocities for three different models.  If
the effect of Doppler boosting is ignored; for example, if the
observed motions are due to the propagation of shocks rather than
actual bulk motion, then most jets will lie close to the plane of the
sky and have apparent speeds, $\beta_\mathrm{app}$ close to
$\beta_\mathrm{p}$. If we take into account the effect of Doppler
boosting, then, in a flux limited sample, with a single Lorentz
factor, most sources are found to lie close to the line of sight, and
have an apparent velocity, $\beta_\mathrm{app} \sim
\beta_\mathrm{p}\gamma$. In the typical situation which we consider,
where $\beta_\mathrm{p} \sim 1$, then in the absence of Doppler
boosting, case most sources appear to have an apparent velocity close
to $c$, whereas in the Doppler boosted case, most sources have an
apparent velocity close to $\gamma c$. Note that in the absence of
Doppler boosting, there are virtually no sources with apparent speeds
less than $\beta c$ since this would require rare end-on orientations,
The randomly oriented model also drops much more sharply with
increasing $\beta$ than the model with a range of Lorentz factors.


\cite{BK79} suggested that one of the observed jet components might be
the stationary feature of the approaching jet at the point where it
becomes optically thin, so that the flux density of this ``core''
component as well as the moving component is Doppler boosted. The
observed overall spectrum is typically flat, which may due to the
superposition of different parts of the jet, which each have different
Doppler factors, causing the synchrotron cutoff to appear at different
frequencies. Consideration of differential Doppler boosting led to the
concept of unified models (e.g., \citealt{OB82,B89,UP95}) which is
usually invoked to understand the observed differences in the
properties of quasars, active galaxies, and BL~Lac objects as due to
the orientation of the relativistic beam and an obscuring torus with
respect to the line of sight. Thus comparison of apparent jet speeds
with optical classification is an effective test of these unified
models and also serves to refine their parameters.

\subsection{\it Jet Kinematics  \label{kinematics}}

The motion statistics collected for our sample address many physical
questions related to relativistic jets. The process of their formation,
that is, the initial acceleration and collimation of the flow, may be
constrained by studying the speeds as a function of distance from the
beginning of the visible jet (often referred to as the ``core''), and also by
comparing the times when flares occur with the back-extrapolated epochs
at which moving features appear to originate. Long-term multi-epoch
observations can show whether there are accelerations or decelerations
farther down the jets, and whether radio features follow straight or
curved trajectories.

In order to discriminate between various jet models, and then to
refine the relevant ones, it is important to establish whether jets
exist as predefined channels, along which multiple moving features can
be seen, or whether instead successive components follow different
trajectories, at either the same or at different speeds. Viable models
of jet formation will also need to be able to reproduce the observed
Lorentz factor distribution. The moving features may in fact be
patterns, caused by the propagation of shocks rather than the flow of
matter.  This can be studied by comparing the Lorentz factors inferred
from the motions to the Lorentz factors estimated by other means, such
as variability, brightness temperature, and (possibly) the presence of
X-ray and gamma-ray emission. By probing for correlations between the
apparent velocities and other quantities such as the radio or X-ray
luminosity, more can be learned about the physical parameters relevant
for jet formation. Finally,
studying the distribution of velocities as a function of optical host
type is relevant to constraining unification models.

Most of the jets in our study are well collimated and are unresolved
transverse to their flow, although there often is significant
curvature. In some radio galaxies with two sided structure, the
appearance of the source is very frequency dependent, suggesting
free-free absorption in a disk or torus, probably associated with the
accretion disk surrounding the central engine.

We found several jet features to have apparent negative velocities;
that is, they appear to be approaching rather than receding from the
core. Most of these apparently negative velocities are consistent,
within the errors, with no significant motion. Observations extending
over a longer time frame are needed to determine if these inward
motions are real. Apparent inward motion may be produced if there is a
newly emerging jet feature that is ejected from the core, and the
combination is not resolved by our beam. This would cause a
shift in the measured position of the centroid and a corresponding decrease with
time in the apparent separation of the core and other jet features. It
is also possible that the true core is not seen, possibly due to
absorption, and that the only observed features are parts of a jet
which are moving with different velocities. If the furthest component
is moving with a slower velocity than the one closest to the obscured
core, then they will appear to be approaching each other. The apparent
decrease in component separation from the core could also be due to
component motion away from the core along a highly curved jet which
bends back toward the line of sight, so that the projected separation
from the core appears to decrease with time. Finally, the moving
features could be only patterns in the flow, some of which might even
be moving backward. \cite{IP96} have discussed an electron-positron jet
model where an observer located close to the direction of motion will
see backward-moving knots. We note that any model involving patterns
must account for the the very large numbers of observed outward
motions, as opposed to inward ones. For this reason, many of the
simplest ``moving marquee'' models for superluminal motion have already
been ruled out. In the case of 0735+178, the most distant feature
appears to be moving inward, but this is probably an artifact of the
complex brightness distribution whose centroid shifts when the
intensity distribution changes. 

\subsubsection{\it Velocity Distributions}

\begin{figure}[t]
\begin{center}
\resizebox{1.0\hsize}{!}{\includegraphics[trim=0cm 0cm 0cm -0.2cm,clip]{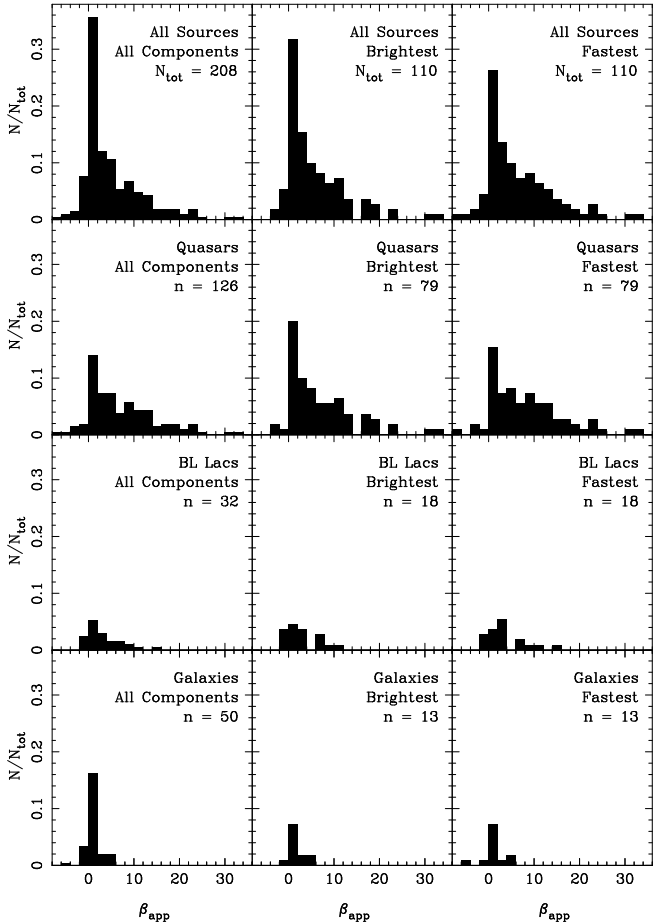}}
\end{center}
\caption{\label{p3histall}
Distribution of the apparent linear velocity in all sources with a
quality code `E' or `G' and which have measured redshifts. The left
hand column displays the distribution for all individual features which
we have observed. Distributions in the center and right columns show
only one feature per source, the brightest or the fastest respectively.
Sources are divided by optical class in the second, third and fourth
rows of the figure.
}
\end{figure}

\begin{figure}[t]
\begin{center}
\resizebox{1.0\hsize}{!}{\includegraphics[trim=0cm 0cm 0cm -0.2cm,clip]{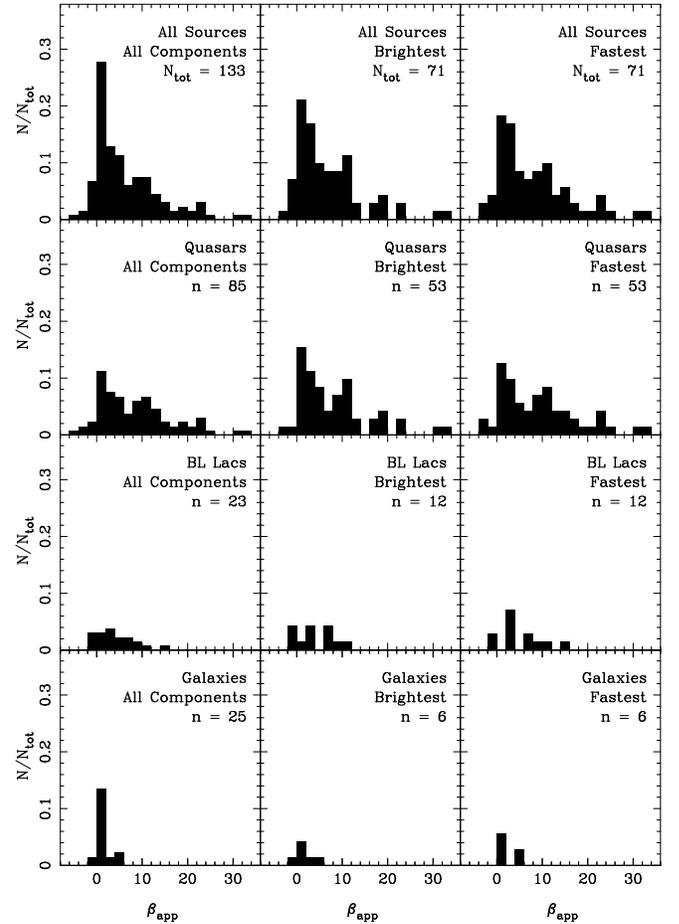}}
\end{center}
\caption{\label{p3histmoj}
Same distributions as in Figure~\ref{p3histall} except only those
sources contained in our representative flux density-limited sub-sample
are included.
}
\end{figure}

Figure~\ref{p3histall} shows the distribution of the observed values of
$\beta_\mathrm{app}=1.58\times 10^{-2}\,\mu\,D_\mathrm{A}\,(1+z)$,
where $D_\mathrm{A}$ is the angular size distance to the radio source
in megaparsecs and the angular velocity, $\mu$, is in mas per year.
Figure~\ref{p3histmoj} shows the same distributions for those jets
found in the representative sub-sample described in
\S~\ref{subsample}.  We include in these figures only those jet
features which we have been able to measure with a quality code of `E'
or `G' and which have measured redshifts.


The central and right-hand columns of Figures~\ref{p3histall} and
\ref{p3histmoj} show the motions of only one feature per source, the
brightest and the fastest respectively. The brightest feature in each
jet is defined as the one with the largest peak flux density at the
epoch for which the source image was presented in Papers I or II,
although in nearly all cases this does not change over the period
covered by these observations. These features generally have
well-determined motions; however, in a few cases, the brightest jet
feature does not have an `E' or `G' quality code, usually because it
has a large diffuse structure. In these cases, we have substituted the
brightest component which did have an `E' or `G' quality code. For the
right-hand panel of Figure~\ref{p3histall}, the fastest feature within
a source is simply defined as the one with the largest (absolute)
linear speed with a quality code of either `E' or `G'.

K--S tests indicate that at a 95 percent confidence level, there is
no difference in the distribution of the jet speeds, independent of
whether we consider the brightest, fastest, or even all the jet
features within a source. We consider the brightest feature of each
source to be the most representative for our analysis, since in the
case of sources with weak secondary features, observations having a
limited sensitivity or dynamic range might only detect the brightest
jet feature.

The distributions shown in Figures~\ref{p3histall} and \ref{p3histmoj}
are also sub-divided according to optical class. We find that the
velocity distributions for the quasars, radio galaxies, and BL~Lacs
are mostly concentrated in the same range ($0<\beta_\mathrm{app} <
15$).  However, the quasars have a tail ranging up to
$\beta_\mathrm{app} \sim 34$, while jets associated with active
galaxies have a narrow range of velocities with
$\beta_\mathrm{app}\leq 6$. A K--S test confirms at the 98 percent
confidence level that the quasars have a different speed distribution
than the galaxies and BL~Lac objects, while the distribution of speeds
for the galaxies and BL Lacs are statistically
indistinguishable. Examination of Figure~\ref{p3histall} and
Figure~\ref{p3histmoj} suggests that the distribution of quasar
velocities may be bi-modal with a minimum near $\beta_\mathrm{app}\sim
5$. We have compared our observed distributions with one and two
Gaussian distributions and find in each case a significantly lower
value of reduced $\chi^2$ for the two component distribution for all
of the quasar distributions.  However, we are reluctant to quantify
this further, as the observed distributions are clearly more complex
than can be represented by two Gaussians.

We have also compared the dispersion in the speed of different
features within each jet, with the dispersion of the average jet
speeds.  We find
that the dispersion of the average jet speeds among the 50 sources
with two or more features with quality factors E or G is 6.3$c$,
whereas in all but four jets (92 percent) the dispersion in the speed of
individual jet features is smaller than this value.  This suggests
that there is an underlying flow, characteristic of each jet, that
determines the speed of individual features with the jet.

\subsubsection{\it Apparent Speed and Apparent Luminosity}

\begin{figure}[thb]
\begin{center}
\resizebox{1.0\hsize}{!}{\includegraphics[trim=0cm 0.5cm 0cm 0.9cm,clip]{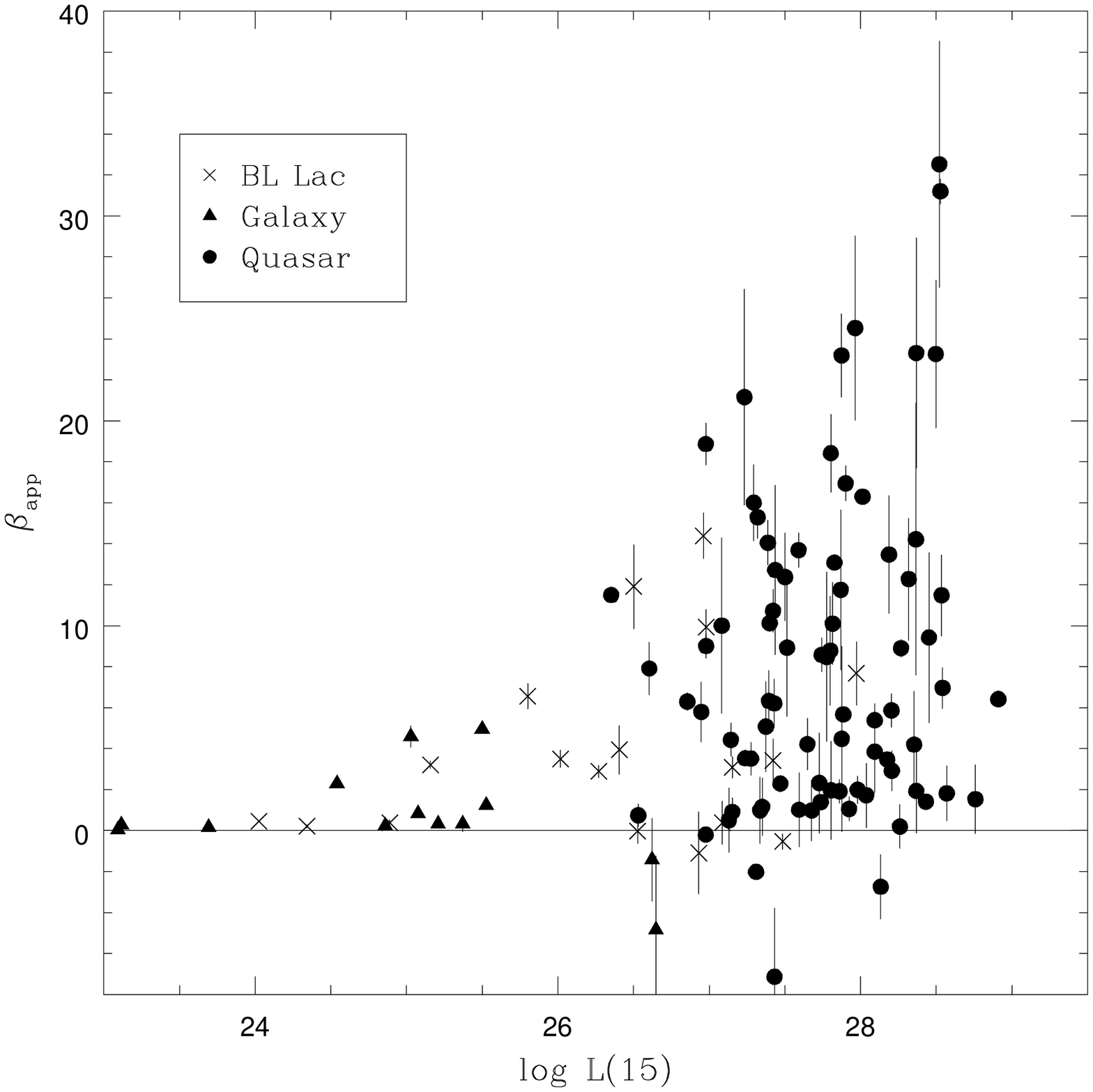}}
\end{center}
\caption{\label{p3betalum}
Apparent speed of the fastest feature found in 110 jets as a
function of the radio luminosity.
}
\end{figure}

Figure~\ref{p3betalum} shows the relation between $\beta_\mathrm{app}$
on apparent VLBA core luminosity at 15~GHz. There is a distinct upper
envelope to the distribution, which is very similar to that found at
5~GHz in the Caltech Jodrell Flat-spectrum survey by \cite{V95}. In
particular, the low-luminosity sources all have slow apparent speeds.
If we divide the sources at the median luminosity of
$L_{15}<10^{27.4}$~W\,Hz$^{-1}$, a K--S test yields less than a 1\%
probability that speed distributions for the brightest components of
the high and low luminosity sources have the same parent distribution.
We note that if the apparent jet speeds that we are measuring are
simply random patterns, then no envelope would be expected in
Figure~\ref{p3betalum}. 

As discussed by \cite{LM97}, the presence of this envelope does not
necessarily imply that {\it intrinsically} faint jets have low {\it
intrinsic} speeds. As discussed in \S~\ref{pgamma}, in practice there is
likely to be a spread in Lorentz factors.  The lowest-luminosity
sources in a flux-limited sample will tend to lie at low redshifts,
where the co-moving volume element is small. Ignoring evolutionary
effects and assuming a constant co-moving space density and a steep
luminosity function, these sources will therefore be representative of
the most common jets in the parent population. The probability of any
of them having {\it both} a high Lorentz factor and small viewing
angle will be very small, especially if the parent population is
dominated by low-Lorentz factor jets (i.e., $n(\gamma) \propto
\gamma^{-1.5}$ as found by
\citealt*{LM97}). This may therefore explain why none of the
low-luminosity sources in Figure~\ref{p3betalum} are highly superluminal. 


\subsubsection{\it Non-Radial and Bent Trajectories \label{non-radial-bent}}

For most of the sources in our sample we have determined only the
radial speed and position angle of the motion with respect to the
assumed core component.  For those sources with sufficiently
high-quality data, we also computed the full two-dimensional vector
proper motion on the sky.  This was accomplished by fitting separately
for the proper motion in both right ascension and declination.  These
results were combined to form a vector velocity, determined by the
speed, $\mu$, and the direction, $\phi$, on the sky relative to the
radio core.  We are particularly interested in comparing the direction
of motion, $\phi$, to the mean structural position angle of the jet,
$<$$\vartheta$$>$.

Table~\ref{non-radial} compiles the vector proper motions for all jet
components in our sample which have at least five epochs of observation
and for which the vector velocity is of at least $5\sigma$
significance. These criteria guarantee that only the highest quality
vector motions are used in our comparison of velocity direction to
structural position angle. Of the 60 component motions that meet these
criteria, we find that 20 are significantly ``non-radial'', meaning the
velocity direction, $\phi$, differs by at least $3\sigma$ from the mean
structural position angle, $<$$\vartheta$$>$. These non-radial motions are
high-lighted in bold in Table~\ref{non-radial}.

Non-radial motions do not follow a direction which extrapolates back to
the jet origin and are by definition considered non-ballistic. The
occurrence of such non-ballistic motions in approximately a third of
the highest quality motions we examined is striking. Assuming the
observed velocities trace the underlying jet flow, this result
indicates that bends in jet direction and/or jet collimation are
common. The actual bends may be small, only a few degrees, but the
observed bends can be large, of the order of $90\arcdeg$, since they
appear amplified by projection.

\begin{figure*}
\begin{center}
\resizebox{0.85\hsize}{!}{\includegraphics[trim=0cm 0cm 0cm 0cm,clip]{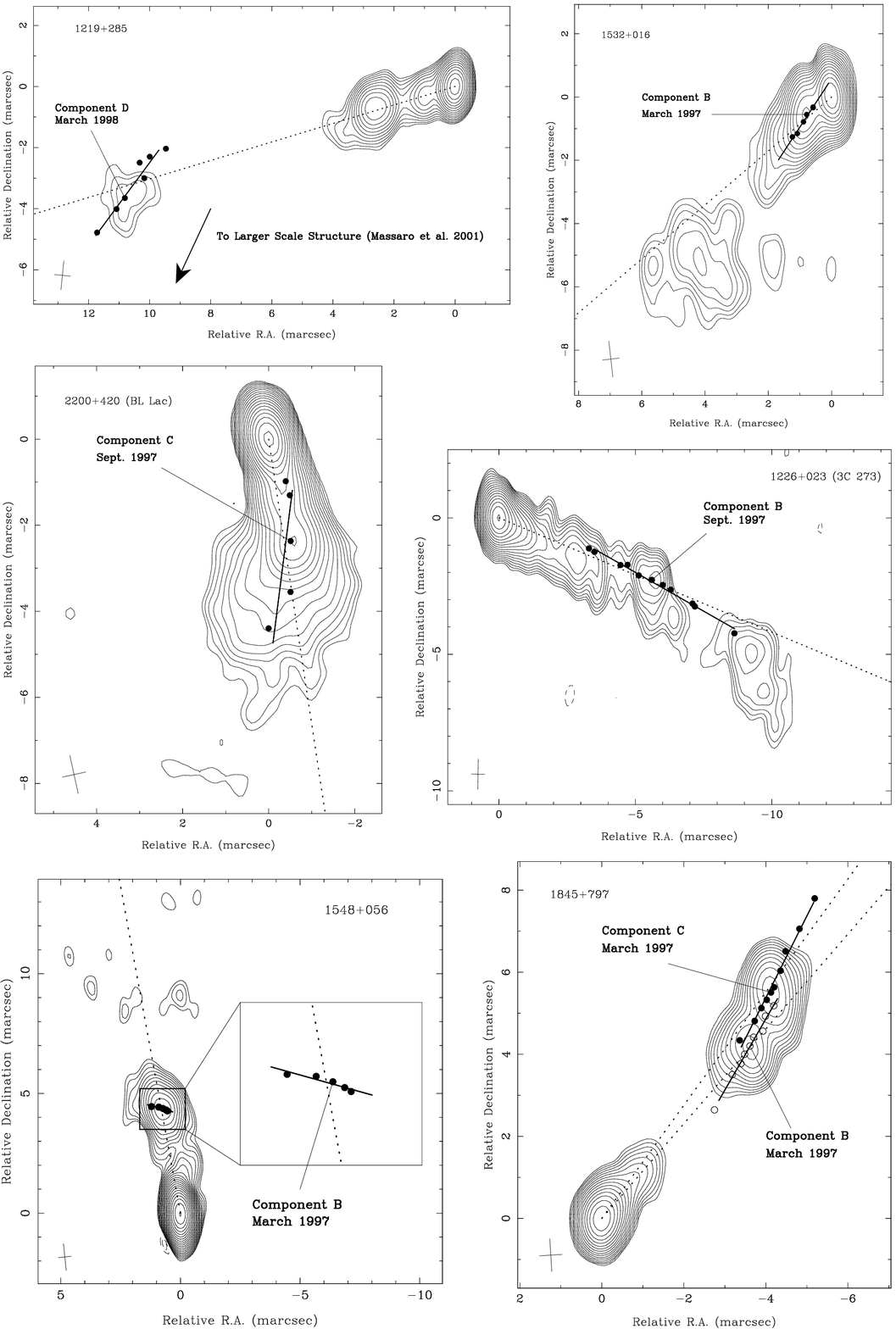}}
\end{center}
\caption{\label{bent}
Selected images of sources with jets that show non-ballistic component
motion. Measured component positions for each epoch are superimposed on
the images along with the vector motion (solid lines) in the RA -- Dec
plane. Dashed lines represent the mean structural position angles,
$<$$\vartheta$$>$, for each component. 
}
\end{figure*}

In Figure~\ref{bent}, we show tracks in the RA -- Dec plane of several
sources that have clearly defined non-radial or bent trajectories. In
general, the trajectories are bent toward the next structure down the
jet, whether seen directly in our maps, or in published larger scale
images. This suggests that jet features may trace out a flow in a
pre-existing curved channel. 1226$+$023 (3C\,273), 1219$+$285,
1532$+$016, and 2200$+$420 (BL~Lac) are excellent examples of this
trend. In 1548$+$056, the component motion appears to be transverse to
the main jet direction which is known from lower resolution images, and
is hinted at in our images, to lie to the north. However, the motion we
observe is directed toward the small extension toward the east, and
looking back toward the core, the jet ridge-line appears to curve into
the component from the opposite direction.


In 3C\,390.3 (1845$+$797) the two most prominent features appear to be
moving with a similar speed of about $2.5c$, but along slightly different
trajectories of $27\arcdeg$ and $30\arcdeg$. Neither trajectory is aligned
with the direction of the core or with the narrow $60\arcsec$ jet that
points toward one of the distant hot spots in position angle $35\arcdeg$
\citep{AWK96}. This is a clear exception to the trend identified above.

In general we do not have sufficient data to robustly detect changes in
velocity, although the high incidence of non-ballistic motion implies
that such changes must occur. However, in BL~Lac, there is evidence
that the moving feature traces the curved ridge-line of the jet. Also,
we have previously reported an abrupt change in both speed and
direction in 3C\,279 that occurred in 1998 \citep{HLK03}.

\subsubsection{\it Two-sided Sources \label{symmetric}}

Seven sources in Table~\ref{motions} display two-sided structure in
which the jet flow appears bi-directional away from a central core. We
have previously reported our results on the two-sided source NGC\,1052
\citep{VRK03}, where the component motion is close to the plane of
the sky and multiple components move in opposite directions from the
core with a velocity, $\beta\sim 0.26$, which is only mildly
relativistic. Plots of component position vs time for all of the two
sided jets are shown in Figure~\ref{cso}. For most of these sources we were
able to identify the central component at each epoch and so determine
the motion of individual features away from the center. In a few cases,
however, such as 1404+286 (epochs 1998.83 and 1999.55) the central core
was not detected at one or more epochs, either because it was weak, or
because the dynamic range was limited for that observation. In these
cases we have interpolated from the positions of the outer features, to
find the ``virtual'' center, and we have used that to determine the
motion of the individual outer features. For the source 1323+321 even
that procedure was not possible since the structure appears to be
two-sided, but we do not see the core at any epoch. Moreover, our
imaging of this source is not completely satisfactory due to its large
angular size and the limited interferometer spacings sampled by our
data.

\subsubsection{\it Peaked Spectrum Sources \label{gps}}

Gigahertz Peaked Spectrum (GPS) sources are characterized by their
sharp low frequency cutoff and general absence of prominent large scale
structure. There is no consensus as to whether the observed spectral
cutoff in these sources is due to synchrotron self absorption or to
free-free absorption \citep{SKC99}. Previous VLBI observations have
suggested that GPS sources have a simple double structure morphology
with little or no significant motions (e.g., \citealt{OD98}). Thirteen
of our sources are classified as GPS sources of which we, so far, have
observations of eight at three or more epochs. Of these, only 1345+125
shows superluminal motion. The mean velocity observed for
the GPS sources in our sample is only ($0.5\pm1.5$)$c$. None of the GPS
sources have $\beta_\mathrm{app}>1.5$.

The existence of a sharp peak in the spectra of these sources implies
that the individual components each peak up at about the same
frequency. This means that the different components probably have
comparable Doppler shifts, and if the spectral cutoff is due to
synchrotron self absorption they should have comparable brightness
temperatures. Some GPS sources, such as 0552+398 and 0642+449 contain
a single unresolved strong component, and so an interpretation in
terms of synchrotron self absorption is required for any reasonable
magnetic field strength. However, for other GPS sources we note the
very different appearance of the core and jet components;
nevertheless, they have a common cutoff frequency. Moreover, many of
the peaked spectrum sources included in our study have sharply bent
jet structure suggesting very different amounts of Doppler boosting
and frequency shift of the self absorption peak. Free-free absorption
probably plays an important role in determining the spectra of these
sources. Since many of the core-jet GPS sources show little or no
motion, we suggest that possibly, the jet flow in these sources is
non-relativistic, or that these sources are seen at very large viewing
angles, rather than that the measured velocity is the advance speed of
a young double lobe radio galaxy as suggested by \cite{OC98}.  Either of
these scenarios would imply a small Doppler shift.  GPS sources do not
have any extended double lobe structure which may be a consequence of
the slow jets which do not carry sufficient energy to form extended
radio lobes.

We note that in some cases, such as CTA\,102 which had a peaked spectrum
in the past, there are large variations in the total flux density
\citep{KKNB02} which probably reflects significant relativistic
boosting. \cite{L03} and \cite{K04} have argued that these sources are
not bona fide GPS sources but ``masquerading blazars'' which often
contain bright transient jet features that dominate their radio
spectrum.

\subsubsection{\it Sources With Extended Double Structure \label{lobedom}}

As a result of our selection criteria, most of the sources included in
our study are dominated by their flat spectrum compact structure; they
are mostly identified with quasars. Some quasars, however, also have
extended structure with relatively steep radio spectra, in addition to
their flat spectrum compact core. We have included a few of these
lobe-dominated sources in our study, although they did not meet our
selection criteria. Extended structure is more common among the active
galaxies. The following sources in Table~\ref{motions} have significant
double-lobe steep-spectrum extended structure: NGC\,315 (0055+300),
NGC\,1052 (0238$-$084), 3C\,111 (0415+379), 3C\,120 (0430+052), M87
(1228+126), 3C\,390.3 (1845+797), and Cygnus~A (1945+405). All are
identified with galaxies containing a relatively strong AGN. All but
NGC\,1052 and 3C\,120 are dominated by the extended steep spectrum
double lobe structure even at 15~GHz. In the framework of standard
relativistic beaming models, radio galaxies and lobe-dominated sources
are presumed to lie close to the plane of the sky, and thus should show
values of $\beta_\mathrm{app}\sim 1$. However, 3C\,111, 3C\,120, and
3C\,390.3 each show rather large apparent motions, and it is not clear
how they fit into standard unification models. 

     These objects are all broad-line radio galaxies (BLRG), although
3C\,120 has also been classified as a narrow-line radio galaxy (NLRG).
The NLRG are thought to lie at rather large angles to the line of sight,
and polarimetry shows that in some cases they contain a central quasar
and a broad line region that are hidden by a dusty torus. The NLRG do
not show strong superluminal motion. 3C\,111, 3C\,120, and 3C\,390.3 do
show superluminal motion and hence must be at small angles to the line
of sight (less than about $11\arcdeg$ for 3C 111). The BLRG are not
well-understood, and they may have a wide range of intrinsic luminosity
and orientation\citep{DBB00}.

On the other hand, the jets of other lobe-dominated sources appear to
be sub-relativistic. The powerful radio galaxy Cygnus~A contains twin
jets pointed toward the distant hot spots. The approaching and
receding jets appear to propagate with velocities of $0.7c$ and $0.2c$
respectively, and we find relatively slow speeds of ($0.16\pm0.03$)$c$
in the parsec scale jet of NGC\,315. In the case of the radio galaxy
NGC\,1052, we have reported a two-sided flow with sub-relativistic jet
speeds of only $0.26c$ \citep{VRK03} on both sides of the core. At
least in these radio galaxies, the intrinsic jet flow close to the
core appears to have only moderate speeds with $\beta_\mathrm{app}\ll
1$.  Thus, it appears that the observed distribution of speeds and/or
degree of jet asymmetry cannot be interpreted simply in terms of the
orientation of a twin relativistic jet with $\beta\sim 1$, unless the
observed speeds do not reflect that actual jet flow speed, or the
radiation from a high speed inner spine is not observed due to a more
slowly moving outer sheath that dominates the emission at large angles
from the jet direction.  As we show in \S~\ref{flux} highly
relativistic jets are relatively rare among the general population of
radio jets.

\subsubsection{\it Comparison With Other Velocity Studies \label{othervels}}

Other recent VLBA monitoring observations (e.g.,
\citealt{JMMW01,JMMA01,BVT01,HOW01,VBT03}) made at other wavelengths
with different resolution and with different sampling intervals are
complementary to our observations and may be used to extend the range
of size and time scales over which jet kinematics may be studied.

\cite{JMMW01,JMMA01} have used the VLBA to study the
motions in a sample of strong sources at 7~mm and 13~mm. These observations probe
the source structure and motions on a scale about three times smaller
than our 2~cm observations, but due to the decreased surface brightness
at 7~mm, they are generally not able to trace the motions beyond a few
milliarcseconds from the core. From the comparison of the observed
motions for the sources in common to our two studies, it is possible to
trace the motions of individual components over a wider range of scales
than is possible from either set of observations alone.

We have compared our velocities determined at 15~GHz with those of
\cite{JMMW01,JMMA01}.  There appears to be little agreement in the
individual source speeds found by the two studies. On average, the
\cite{JMMW01,JMMA01} speeds appear systematically higher, possibly
because their observations were carried out at shorter wavelengths and
probed jet regions closer to the core.

\cite{BVT99,BVT01} and \cite{VBT03} have discussed jet motions in a
large sample of sources at 6~cm, taken from the Caltech-Jodrell CJ
surveys covering declinations greater than $+35\arcdeg$. These
observations have lower angular resolution than ours, and are thus more
sensitive to the lower surface brightness structure located downstream.
\cite{VBT03} quote a mean velocity for quasar and galaxy jets of $2.9c$
and $0.9c$ respectively. This appears be less than the values of
($7.3\pm0.8$)$c$ and ($1.7\pm0.8$)$c$ which we measure at 2~cm for the
brightest features in each source. On the other hand, \cite{JMMW01},
working primarily at 0.7 and 1.3~cm, find systematically faster
velocities in those sources where our samples overlap. These results
suggest that there is a systematic decrease in $\beta_\mathrm{app}$
with increasing wavelength, probably because the observations at
different wavelengths, sample different parts of jet structure. In a
separate paper (R.~C.\ Vermeulen et al., in preparation) we will
discuss, in more detail, the motions of those sources in common to the
6 and 2~cm samples.

\subsection{\it Implications for Relativistic Beaming Models \label{beamingmodels}}
\subsubsection{\it Ballistic Models \label{pgamma}}

The simplest model to consider is a pure ballistic model with a common
flow velocity for all sources, with
$\beta_\mathrm{p}=\beta_\mathrm{b}\sim 1$; and a flow which is
intrinsically symmetric. In this simple case, observations of the
apparent velocities, and the ratio of flux densities of approaching
and receding components, can, in principle, be used to solve uniquely
for $\gamma$ and $\theta$ and thus provide a test of the hypothesis
that $\beta_\mathrm{p}=\beta_\mathrm{b}$. However even early VLBI data
indicated that this simple model is not tenable (\citealt{VC94,LM97}).

In a flux density-limited sample, the combined effect of available
solid angle and Doppler boosting leads to a distribution of
$\beta_\mathrm{app}$ with many values close to $\gamma$ \citep{VC94}.
On the other hand, if Doppler boosting is not an important selection
mechanism, then in most cases $\beta_\mathrm{app}\sim$ 1.
Figure~\ref{rene} shows the expected distribution of apparent
velocities for the case where $\gamma_\mathrm{p} = \gamma_\mathrm{b}$ (see
\S~\ref{beaming}) along with the corresponding distribution for a
randomly oriented sample. None of the distributions shown in
Figure~\ref{p3histall} or Figure~\ref{p3histmoj} are consistent with
either of these simple ballistic models as there is neither evidence
for the peak at $\beta_\mathrm{app}\sim\gamma$ characteristic of the
simple Doppler boosted models, nor for the sharp low speed cutoff at
$\beta_\mathrm{app}\sim 1$ characteristic of models which ignore
Doppler boosting.

How do we reconcile the difference between the observed and predicted
distributions? Either there must be spread in intrinsic velocity (e.g.,
\citealt{LM97}) or a difference between the bulk flow
velocity and the pattern velocity, so that there is less Doppler bias
in favor of observing beams which are oriented close to the critical
angle \citep{VC94}. In Figure~\ref{rene}, we plot a model with a spread
of intrinsic velocity such that $n(\gamma) \propto \gamma^{-1.5}$. This
model has equal pattern and bulk velocities, and reproduces the general
characteristics of the observed distribution of superluminal speeds.
A more detailed analysis will be discussed by M.~L.\ Lister et al.~(in
preparation).

\subsubsection{\it Randomly Oriented Samples}

Based on earlier observations of apparent velocity distributions
\cite{EL90} suggested that the simple model with no Doppler
bias, provided an adequate fit to the observed distributions of
$\beta_\mathrm{app}$ with only a slight adjustment needed which was
satisfied by introducing an obscuring torus. However, comparison of
our data with models which do not include Doppler boosting shows poor
agreement, since, as shown in Figure~\ref{rene}, in the absence of
Doppler boosting, most sources are expected to lie close to the plane
of the sky where $\beta_\mathrm{app}\sim 1$. This appears to be
inconsistent with the tail of the velocity distribution which we find
extending toward larger apparent velocities.  Following the discussion
of \cite{C90} a detailed analysis of the observed velocities shows
that the probability of the sources having been picked at random from
a parent population that is isotropically distributed is less than
$10^{-5}$. We therefore conclude that the sources are preferentially
aligned along the line of sight, as would be expected if Doppler
boosting is in fact important.

\subsection{\it Comparison with Other Relativistic Velocity Indicators} \label{comparison}

While the actual observation of superluminal motion in radio source
jets remains the most direct way of establishing the existence of
relativistic motion, comparison of the observed values of
$\beta_\mathrm{app}$ with other velocity indicators also provides
important tests of beaming models and specifically the relation
between pattern and bulk flow speeds. This includes flux density
variability (\citealt{HAA92, LV99, LVW99}), maximum brightness
temperature (\citealt{KP69, R94, GD96}), the ratio of core-to-extended
radio luminosity (\citealt{OB82}), the gamma-ray luminosity
(\citealt{VM95, HBB99, MSM97, MHR01}), or observations of inverse
Compton X-rays (\citealt{GPC93}). Relativistic boosting also affects
the radio source counts and luminosity functions (\citealt{PU92,
WJ97}) which provide a consistency check on beaming models.
 
\subsubsection{\it Relation Between Observed Velocity and Flux Density
Variations \label{flux}}

Flux density changes are commonly seen in superluminal sources, and
their short time scale is generally taken to imply high brightness
temperatures. The variability time scale, and the time scale for
apparent transverse motion, are both compressed due to the forward
motion, and hence we might expect to see a relation between
$\beta_\mathrm{app}$ and the flux density variability provided that
$\beta_\mathrm{b}$ is related to $\beta_\mathrm{p}$. \cite{LV99} used
variability data at 1.3~cm and 8~mm from the Mets\"ahovi Observatory
to calculate a variability Doppler factor, $\delta_\mathrm{var}$,
assuming an intrinsic brightness temperature characteristic of a
self-absorbed synchrotron source in which the particle and magnetic
energy are in equilibrium. We have recalculated their values using the
cosmology given in \S~\ref{intro} and for different values of
intrinsic brightness temperature. Since $\delta_\mathrm{var}$ varies
inversely as the cube root of the assumed intrinsic temperature in the
synchrotron plasma, $T_\mathrm{b}^\mathrm{int}$, $\delta_\mathrm{var}$
is relatively insensitive to the assumed value of
$T_\mathrm{b}^\mathrm{int}$.


\begin{figure}[t]
\begin{center}
\resizebox{1.0\hsize}{!}{\includegraphics[trim=3cm 4.5cm 4.0cm 6.25cm,clip]{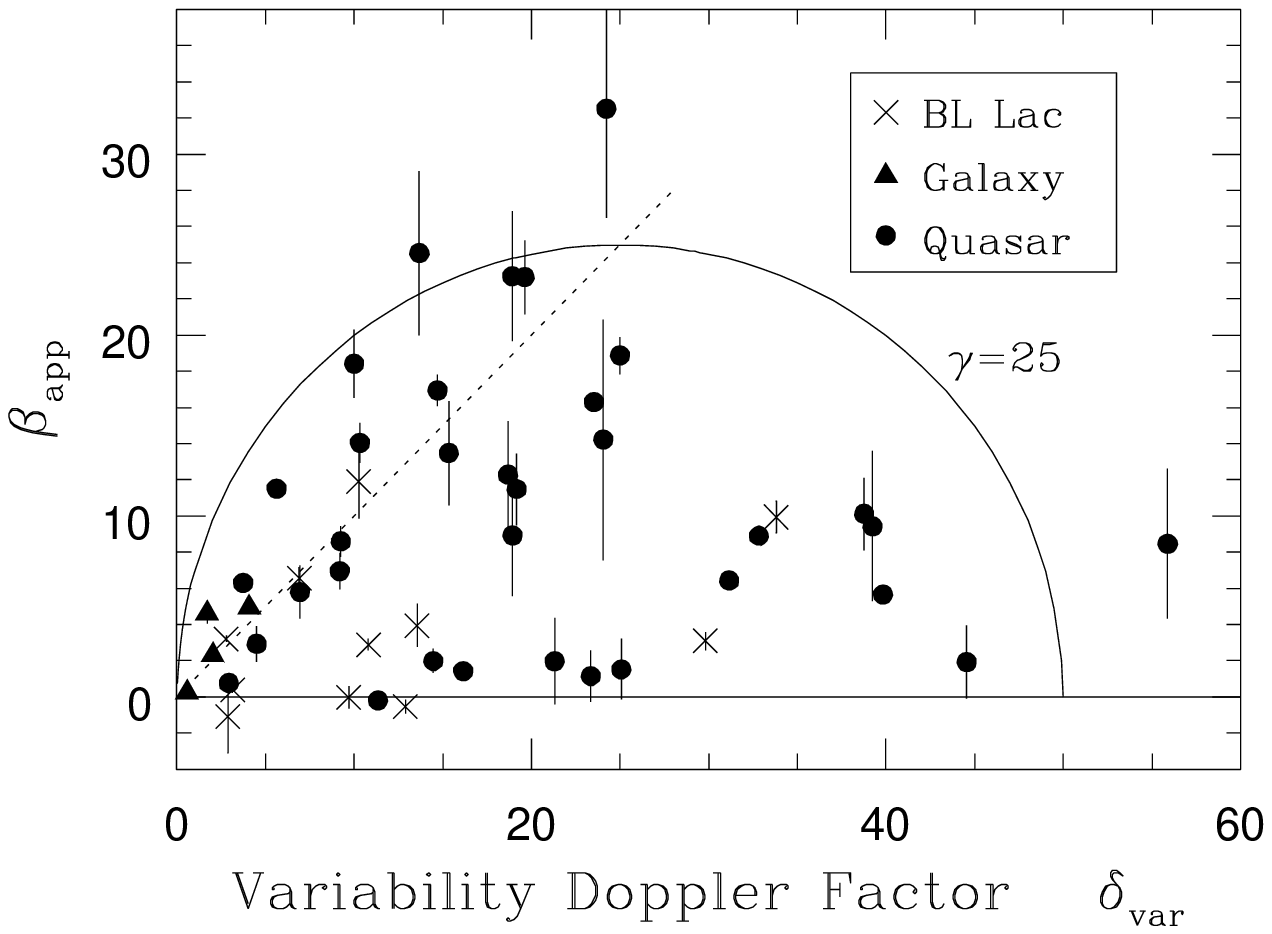}}
\end{center}
\caption{\label{cohen}
Apparent velocity, $\beta_\mathrm{app}$, plotted against variability
Doppler factor, $\delta_\mathrm{var}$, for the fastest component found
in 49 sources calculated using the method of \cite{LV99} with an
intrinsic brightness temperature $T_\mathrm{b}^\mathrm{int}=2\times
10^{10}$~K. The solid line shows the
expected locus of points for Lorentz factor values of 25. The dotted
line represents the $1/\gamma$ cone where
$\delta_\mathrm{var}=\beta_\mathrm{app}$.
}
\end{figure}


In Figure~\ref{cohen}, we plot $\beta_\mathrm{app}$ against
$\delta_\mathrm{var}$ for the 49 sources in common to the Mets{\"a}hovi
and VLBA samples. We calculate $\delta_\mathrm{var}
=[T_\mathrm{b}^\mathrm{var}/T_\mathrm{b}^\mathrm{int}]^{1/3}$ assuming
a spectral index of zero and where $T_\mathrm{b}^\mathrm{var}$ is the
apparent brightness temperature calculated from the variability time
scale, by assuming that it is limited by the size of the source divided
by the speed of light. Five sources which only have components located
at a bend in the jet were excluded, as they probably reflect a standing
shock wave, or perhaps a stationary location in a helical jet where the
flow is closest to the line of sight, and hence is boosted most
strongly. In either case the measured velocity is a poor indicator of
the flow velocity, and not relevant in discussing relativistic effects.
The sample we use contains 5 active galaxies, 14 BL Lacs, and 30
quasars. Most points lie inside the ``$1/\gamma$ cone''
($\beta_\mathrm{app}=\delta_\mathrm{var}$) as they should for a
flux desnity limited sample \citep{VC94}.

\cite{CRH03} have compared Figure~\ref{cohen} with the results of a
simulation generated by randomly picking a flux-limited sample from an
isotropically distributed population with power law distributions of
luminosity and $\gamma$ and for several values of
$T_\mathrm{b}^\mathrm{int}$ over the range $4\times 10^{9}$ to $1
\times 10^{11}$~K. $T_\mathrm{b}^\mathrm{int}\sim 2\times10^{10}$~K
gives the best fit between the measured and simulated data. However,
although there does appear to be an upper limit to $\beta_\mathrm{app}$
that is close to the expected locus for components with $\gamma=25$, the
detailed distribution is not well-matched to that expected from the
simulation \citep{LM97}.

Calculations of the variability Doppler factor using values of
$T_\mathrm{b}^\mathrm{int}$ closer to the inverse Compton limit, $\sim
10^{12}$~K, lead to distributions on the
$\beta_\mathrm{app}\,$--$\,\delta_\mathrm{var}$ plane that are very
different from those expected from the simulations. We conclude that
$T_\mathrm{b}^\mathrm{int}$ is perhaps an order of magnitude below the
inverse Compton limit.

For this application, we have used the \textsl{fastest} feature for
each source, on the grounds that these velocities should be more
representative of the true flow velocities. Slower-moving components,
especially those located at a bend in the jet, may be dominated by
standing shock waves. Forward shock waves might also exist, and trying
to understand their role is a goal of our study. Other geometries have
been suggested for the jet, including a fast ``spine'' which we would
preferentially see, surrounded by a slower shell. In this case, the
spine would probably also control the flux variations, so that using
the fastest (spine) velocity for $\beta_\mathrm{app}$ is appropriate.

We have also examined values of $\delta_\mathrm{var}$ calculated from
the \mbox{UMRAO} data base at our wavelength of 2~cm, and noted a large
dispersion between Doppler factors deduced from these data and
the shorter wavelength Mets{\"a}hovi data. Thus, the robustness of the
Doppler factors calculated in this way appears to be uncertain. The
Michigan data cover a longer time span, but are at longer wavelengths
where individual outbursts appear to overlap in time. 


\begin{figure}[t]
\begin{center}
\resizebox{1.0\hsize}{!}{\includegraphics[trim=1.5cm 1.5cm 5.5cm 10.85cm,clip]{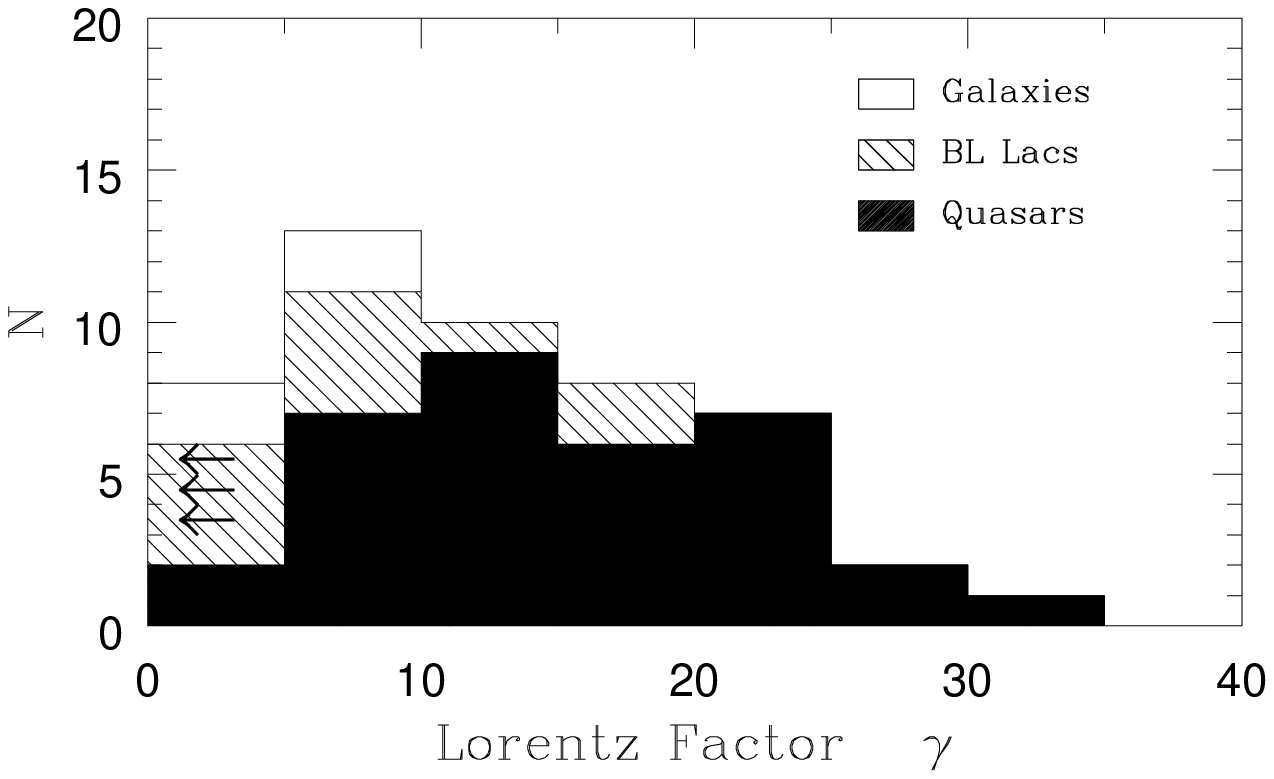}}
\end{center}
\caption{\label{cohenhist}
Histogram of Lorentz factors for 49 sources, calculated from
$\beta_\mathrm{app}$ and $\delta_\mathrm{var}$ based on
$T_\mathrm{b}^\mathrm{int}=2\times 10^{10}$~K.
}
\end{figure}


The Lorentz factor, $\gamma$, in a superluminal jet is important,
because it is intrinsic to the jet, whereas $\beta_\mathrm{app}$ and
$\delta_\mathrm{var}$ are the observables that depend on the jet
orientation. The distribution of $\gamma$ may give information on the
physics of the collimation region. From the values of
$\beta_\mathrm{app}$ and $\delta$ shown in Figure~\ref{cohen} we
calculate the actual Lorentz factors, shown as a histogram in
Figure~\ref{cohenhist}. We believe that the apparent deficit in the
first bin of Figure~\ref{cohenhist} for the quasars,
may be a selection effect or just due to small number statistics. To be
fully consistent with the analysis above, and the choice
$T_\mathrm{b}^\mathrm{int}=2\times 10^{10}$~K, the Lorentz factors
should have a power-law distribution. However, the number of objects is
too small, and the errors are too uncertain, to make a meaningful
comparison. A small Lorentz factor implies a low velocity and small
flux density variations, and the latter especially is less likely to be
measurable. However, the calculated Lorentz factors are useful in
showing that there must be a wide range of $\gamma$ in the superluminal
sources. The galaxies all have rather low Lorentz factors, while the
quasars have a broad distribution up to $\gamma\approx 30$.

\subsubsection{\it Gamma-Ray Sources \label{gamma}}

Many of the sources included in our study have been cataloged as
strong gamma-ray sources according to measurements made by the EGRET
detector on board the {\it Compton Gamma-Ray Observatory} 
\citep{VM95, HBB99, MSM97, MHR01,SRM03}.  It is generally thought that the
gamma-ray emission occurs deep within the relativistic jet. Arguments
based on size limits deduced from time variability and the cross
section for pair production suggest that the gamma-ray emission, like
the radio emission, is Doppler boosted \citep{DS94}. In fact, the
gamma-rays may be even more strongly beamed than the radio emission
since the former generally have a steeper spectral index, $\alpha$, and
the flux density boosting varies as $\delta^{2-\alpha}$ for continuous
jets. Also, \cite{Dermer95} has shown that if the bulk of the
gamma-rays are produced by external Compton scattering off photons
associated with the accretion disk, the resulting gamma-ray emission
will be boosted by an additional factor of $\delta^{1-\alpha}$.

If gamma-ray loud AGN do indeed have systematically high Doppler
factors, then we might expect to see a different apparent speed
distribution for them than for AGN that have not been detected in
gamma-rays. The situation is complicated by the possibility that the
gamma-ray loud jets may be seen inside the critical angle for maximizing
superluminal motion, ($1/\gamma$ radians), and therefore might have slow
apparent projected speeds. However, Monte Carlo simulations based on a
simple linear relationship between radio and gamma-ray luminosity (e.g.,
\citealt{LM99,L99}) confirm that in a flux density limited radio sample,
AGN detected by EGRET should have typically higher speeds than those
that were not detected in gamma-rays.

\cite{JMMW01} recently measured the apparent speeds of 33
EGRET-detected AGN. They found a mean value of $16c$ for the fastest
component in each source and concluded that the gamma-ray sources have
larger Lorentz factors than the general population of radio sources.
However, \cite{JMMW01} did not have a non-gamma-ray control sample
observed in the same way and at the same wavelength with which to
compare their results.
 
\begin{figure}[t]
\begin{center}
\resizebox{1.0\hsize}{!}{\includegraphics[trim=0cm 0cm 0cm -0.1cm, angle=0,clip]{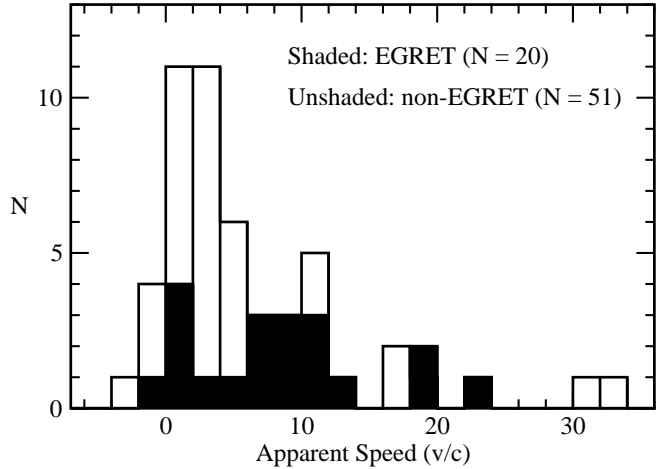}}
\end{center}
\caption{\label{egret}
Histogram of the brightest component speed in EGRET and
Non-{\it{}EGRET} detected sources for our representative flux
density limited sample.}
\end{figure}

We have classified our sample into EGRET and non-EGRET
sources based on the list of \cite{MHR01} and recently modified by
\cite{SRM03}. These authors classify gamma-ray sources from the third
EGRET catalog \citep{HBB99} as ``highly probable'' and
``plausible'' AGN identifications, based on Bayesian statistics and
their proximity to bright flat-spectrum radio sources. In
Figure~\ref{egret} we show the distribution of measured speeds for the
EGRET detected and non-detected sources. For the brightest jet
component found in the 18 EGRET sources in our representative
sample we found a median speed of ($8.0\pm1.6$)$c$ compared with a
value of ($3.9\pm1.1$)$c$ for the 53 sources with no EGRET
detections. A K--S test suggests that the difference appears
significant at the 90 percent level. For the purpose of this analysis
we have included the two ``plausible'' EGRET sources, 2230+114
and 1156+295 as detections. Re-classifying them as non-detections
sources had no effect on our results, and neither did excluding the
sources with negative velocities from our analysis.


These results are consistent with the idea that the radio emission from
gamma-ray quasars is indeed more strongly beamed than for the whole
radio quasar population. However, our samples are incomplete and may
therefore be biased.  


\subsection{\it The Angular Velocity Redshift Relation \label{muz}}

\begin{figure}[t]
\begin{center}
\resizebox{1.0\hsize}{!}{\includegraphics[trim=2cm 6.5cm 4.0cm 2.85cm,clip]{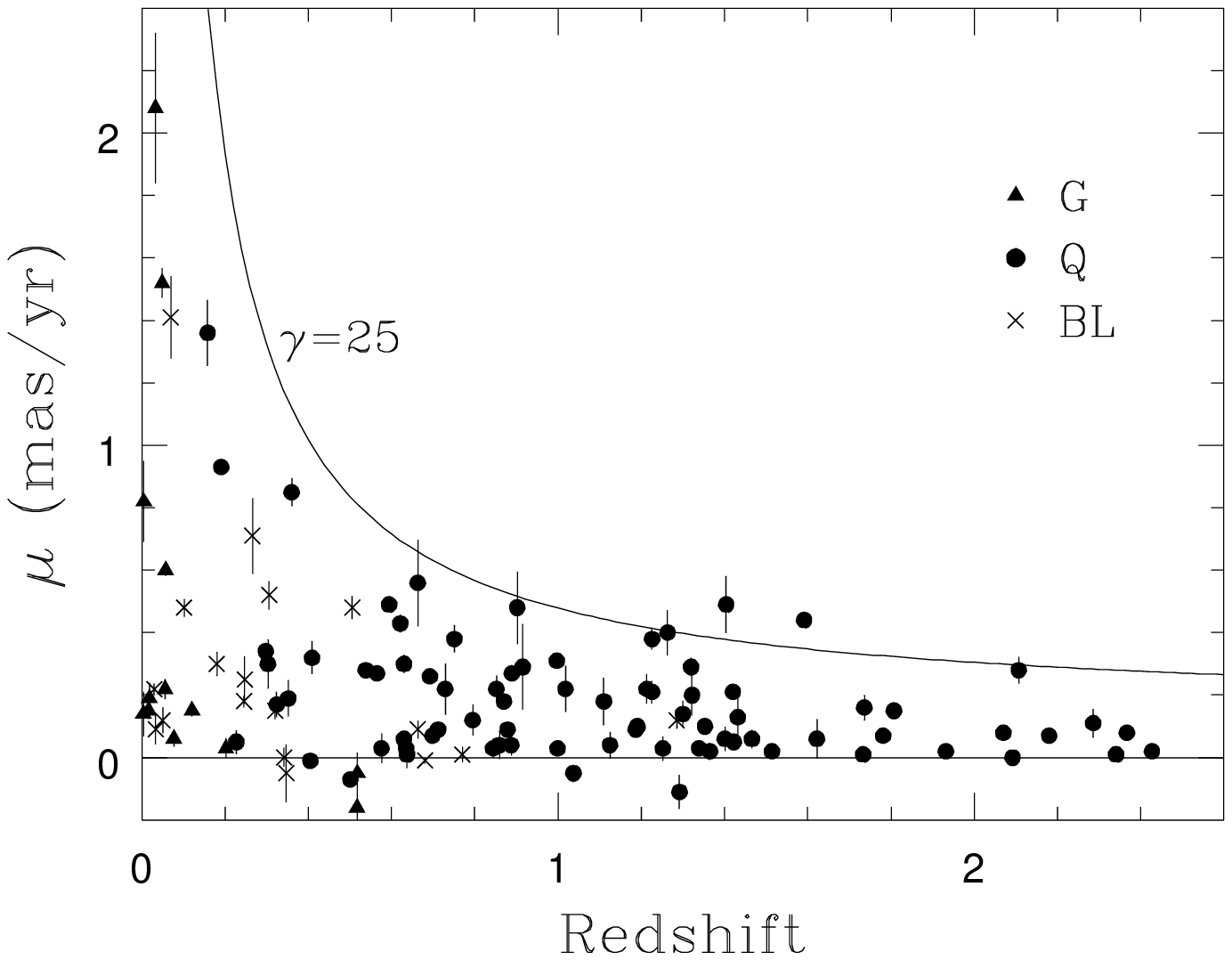}}
\end{center}
\caption{\label{muzfig}
$\mu$--$z$ diagram showing the distribution of angular velocity for the
fastest component of the 110 sources which has a quality factor of `G' or `E'.
The solid line is the maximum value of $\beta_\mathrm{app}$ for
$\gamma = 25$.}
\end{figure}

Figure~\ref{muzfig} shows the measured values of angular velocity, $\mu$,
vs redshift, for the fastest `E' or `G' rated component found in the
sources in Table~\ref{motions}. The line represents $\mu_\mathrm{max}$,
the fastest proper motion a source can display if it has $\gamma = 25$.
The variables are the observables, uncontaminated by modeling, and
hence are of value in showing directly that (a) at all redshifts, the
observed velocities are not clustered near the maximum value as
expected from the simple ballistic models with a single Lorentz factor
for all sources, (b) low values of $\mu$ are seen at all $z$, and (c)
high values of $\mu$ are seen only at low $z$. This is true for
galaxies, BL~Lacs, and quasars and appears inconsistent with
non-cosmological interpretation of quasar redshifts as proposed by
\cite{GRB04}.


An early version of this plot \citep{CBP88} was used to show that the
standard paradigm for superluminal motion provided a crude upper bound to
the points in Figure~\ref{muzfig}, and that the maximum value of
$\beta_\mathrm{app}$ in 32 sources was about 13 ($H_0=70$, $q_0=0.5$).
The $\mu_\mathrm{max}$ line in Figure~\ref{muzfig} is, similarly, a
crude upper bound to the measured points, and shows that
$\beta_\mathrm{app,max}\sim 25$. Note that the cosmology used in this
paper makes $\beta_\mathrm{app}$ somewhat larger than for a cosmology
with $\Omega_\mathrm{m}=1$ and $\Omega_\Lambda=0$.

The data in \cite{CBP88} were compiled at assorted frequencies,
mostly below 15~GHz, and corresponded to measurements made at
different distances from the core which may explain the difference in
$\beta_\mathrm{app,max}$.  As we have discussed in \S~\ref{othervels}
there is some evidence that observations at high frequencies give
higher values of $\mu$ than lower frequencies.

\cite{VC94} and \cite{LM97} have shown that, if Doppler boosting is
important, then even rather small samples of superluminal sources will
show $\beta_\mathrm{app,max}$ near the maximum value of the distribution
of $\gamma$, $\gamma_\mathrm{max}$. Hence we expect that
$\gamma_\mathrm{max}\sim 25$ for the sample shown in
Figure~\ref{muzfig}. This is marginally consistent with
Figure~\ref{cohenhist}, reflecting that Lorentz factors calculated
according to the method in \S~\ref{flux} have rather large
uncertainties.

\section{SUMMARY}
\label{summary}

We have studied the kinematics of a large well-defined sample of 110
quasar and active galaxy jets and find a distribution of apparent
velocities typically between 0 and $14c$ but ranging up to about
$34c$.  There is evidence for a characteristic velocity in each jet
which may represent the true plasma flow velocity. We have found that
quasar jets generally have larger apparent velocities than jets
associated with BL~Lac objects and active galaxies, although the
distributions overlap.  Our measured values of $\beta_\mathrm{app}$
are consistent with the Doppler factors, $\delta_\mathrm{var}$,
calculated from time variability and a parent population having a
steep power law distribution of intrinsic Lorentz factors extending
down to moderate velocities and an intrinsic brightness temperature
close to $2\times10^{10}$~K. This is close to the value expected if
the particle and magnetic energy densities in the jet are comparable
(\citealt{R94,SG85})

In approximately one-third of the well
studied jets, we find evidence for non-ballistic motion; that is the
flow is not along the direction away from the core. In most of the
jets, we find no deviation from a constant speed; but in a few
sources, we do see evidence for changes in speed and direction of
individual features.  Mostly, the flow appears to lie along the
direction of the local jet orientation. However, in some cases the
flow has a significant non-radial component, which points toward more
distant parts of the jet. This suggests that there is a continuous
flow along a pre-existing channel. Contrary to the assumption of the
simple unified models, in some jets, the intrinsic flow appears to be
with speeds much less than $c$.

Observations made at higher frequencies sample jet features located
closer to the core, and they typically show larger apparent velocities
than we observe at 15~GHz, while lower frequency observations show yet
smaller speeds. Sources with stable GPS spectra show little or no
motion; the jet flow in these sources may be non-relativistic or lie
in the plane of the sky with a
correspondingly small Doppler shift.

We find that jets of quasars which have been observed as strong
gamma-ray sources have marginally higher speeds than those which are
not gamma-ray sources. This is consistent with models where the
gamma-ray sources have more highly relativistic jets and are aligned
closer to the line of sight. However, our analysis is limited by both
small number statistics, and the uncertainties in the ever changing
analysis of the EGRET catalogs. Also, with the limited range of
flux density observed by EGRET, there is no well defined class of
gamma-ray loud and gamma-ray quiet sources analogous to the radio loud
and radio quiet classifications. More sensitive observations with the
next generation of gamma-ray observatories, such as {\it GLAST},
combined with jet speed data for our complete radio sample of 133 radio
sources, should be very useful for investigating gamma-ray production
mechanisms in AGN jets and relating the gamma-ray properties to the
observed jet outflow.

E.\ Ros et al.~(in preparation) have extended these observations and
analysis through 2001 and 2002. M.~L.\ Lister et al.~(in preparation)
have defined a more complete sample of 133 sources and are continuing
the observations of these sources including linear and circular
polarization (D.~C.\ Homan et al., in preparation). Observations with
this new sample will allow a more robust comparison with models, a
better estimate of the distribution of intrinsic Lorentz factors, and a
start to understanding the evolution of jet magnetic fields.

\acknowledgements

The VLBA is a facility of the National Radio Astronomy Observatory
which is operated by Associated Universities, Inc., under a
cooperative agreement with the National Science Foundation. We thank
Hugh and Margo Aller, and Tigran Arshakian for many valuable
discussions, Mike Russo, Andrew West, and John Armstrong for their
help with the data analysis, and the NRAO staff for their support in
the data acquisition and correlation. We also thank the referee for
his very constructive suggestions which have helped to clarify our
presentation.  Ed Fomalont and Leonid Gurvits kindly allowed us to use
their 1998 and 1999 images to supplement our own observations during
that period.  We have made use of the data base from the UMRAO which
is supported by funds from the University of Michigan and additional
VLBA observations of Cygnus A by Uwe Bach to help interpret our own
data for this source.  Part of this work was done while RCV held an
appointment at Caltech.  KIK thanks Caltech and the MPIfR for their
support and hospitality during several visits.

%
%



\clearpage
\LongTables


\begin{deluxetable*}{llcclccrc} 
\tablecolumns{9} 
\tabletypesize{\scriptsize} 
\tablewidth{0pt}  
\tablecaption{\label{sample}General Source Properties}  
\tablehead{\colhead{IAU} & \colhead {} &  
\colhead{Sub-sample} & \colhead{Optical} & \colhead{ } &  
\colhead{Radio} &\colhead{Radio} &  \colhead{$S_{VLBI}$} &\colhead{EGRET} \\ 
\colhead{Name} & \colhead {Alias} &  
\colhead{Member} & \colhead{Class} & \colhead{Redshift} &  
\colhead{Spectrum} &\colhead{Morphology} & \colhead{(Jy)}\ &  \colhead{ID} \\ 
\colhead{(1)} & \colhead{(2)} & \colhead{(3)} & \colhead{(4)} &  
\colhead{(5)} & \colhead{(6)} & \colhead{(7)} & \colhead{(8)} & \colhead{(9)} } 
\startdata 
0003$-$066  & NRAO 005  & Y & B& 0.347 & Flat & CJ &   2.84 & N    \\*
0007+106  & III Zw 2, Mrk 1501  & Y & G& 0.0893 & Flat & N &   1.29 & N    \\*
0016+731  &   & Y & Q& 1.781 & Flat & CJ &   0.98 & N    \\*
0026+346  &   & N & G\tablenotemark{a}& 0.517 & Flat & CJ &   0.65 & N    \\*
0035+413  &   & N & Q& 1.353 & Flat & CJ &   0.53 & N    \\*
0048$-$097  &   & Y & B& \n & Flat & N &   1.65 & N    \\*
0055+300  & NGC 315  & N & G& 0.0165 & Flat & CJ &   0.82 & N    \\*
0106+013  &   & Y & Q& 2.107 & Flat  & CJ &   2.97 & N    \\
\enddata 
\tablecomments{Columns are as follows: (1) IAU Name (B1950.0); (2) Other Name; (3) Indicator for the flux density limited sub-sample (see \S~\ref{sampledef}); (4) Optical classification according to the \cite{VCV01} catalog, where Q = quasar, B = BL Lac object, G = active galaxy, and U = unidentified; (5) Redshift from \cite{VCV01} (the redshifts for 0026+346, 0727$-$115 and 1155+251 are from \citealt*{Z02}, that of 0202+149 is from \citealt*{SRK96}, 0218+357 is from \citealt*{Law96}, 0754+100 is from \citealt*{CFK03} and that of 2209+236 is from \citealt{SRM03}); (6) Description of radio spectrum (see \S~\ref{sampledef}); (7) Radio morphology classification, where CJ= core-jet, 2S = two-sided jet, N = naked core, LD = lobe-dominated, and CSS = compact-steep spectrum object; (8) Strongest 2 cm total cleaned VLBA flux density at any epoch, in Jy; (9) EGRET gamma-ray source identification according to \cite{MHR01} and \cite{SRM03}, where Y = highly probable identification, P = probable identification, N = no identification.}

\tablenotetext{a}{Source not in the \cite{VCV01} catalog.}
\tablenotetext{b}{Source classified as quasar in the \cite{VCV01} catalog.}
\tablenotetext{c}{Source classified as galaxy in the \cite{VCV01} catalog.}
\tablenotetext{d}{The source 0218+357 is a gravitationally lensed AGN which we do not include in our statistical analysis.}
\tablenotetext{e}{The complex source 0316+413 (3C~84)  satisfies our selection criteria, but was observed as part of a separate monitoring program \citep{WDR00}. It is included in our statistical analysis for completeness.}

\tablenotetext{f}{Source has large or diffuse structures that are not adequately sampled by our observations.}

\end{deluxetable*} 


\begin{deluxetable*}{ccrrrrrccc}
\tablecolumns{10}
\tablewidth{0pc}
\tabletypesize{\scriptsize}
\tablecaption{Proper motions.\label{motions}}
\tablehead{\colhead{Object} & \colhead{Comp.} & \colhead{$N$} & 
\colhead{$<R>$} & \colhead{$<\vartheta>$} & \colhead{$\mu_r$} &
\colhead{$\beta_\mathrm{app}$} & \colhead{$t_0$} & \colhead{Distinct component} & 
\colhead{Rating}\\  
&&& \colhead{(mas)} & \colhead{(deg)} & \colhead{(mas/yr)} && 
\colhead{(years)} &  & \\
\colhead{(1)} & \colhead{(2)} & \colhead{(3)} & \colhead{(4)} & 
\colhead{(5)} & \colhead{(6)} & \colhead{(7)} & \colhead{(8)} & 
\colhead{(9)} & \colhead{(10)} }
\startdata 
0003$-$066\tablenotemark{$\star$}
 & B\tablenotemark{\dag} & $5$ & $1.1$ & $-72$ & $-0.05\pm0.09$ & $-1.1\pm2.0$ & \nodata & Y & G\\ 
 & C & $4$ & $3.0$ & $-77$ & $0.01\pm0.02$ & $0.3\pm0.4$ & \nodata & Y & E\\ 
 & D & $5$ & $6.2$ & $-78$ & $-0.02\pm0.06$ & $-0.3\pm1.3$ & \nodata & Y & E\\ 
0016$+$731\tablenotemark{$\star$}
 & B\tablenotemark{\dag} & $4$ & $0.8$ & $-50$ & $0.074\pm0.003$ & $5.7\pm0.2$ & $1987.2\pm0.5$ & N & G\\ 
0026$+$346
 & B & $5$ & $10.4$ & $-124$ & $0.03\pm0.23$ & $0.9\pm7.0$ & \nodata & N & P\\ 
 & C\tablenotemark{\dag} & $5$ & $12.2$ & $-129$ & $0.15\pm0.12$ & $4.7\pm3.8$ & \nodata & Y & G\\ 
 & E & $5$ & $14.7$ & $56$ & $-0.16\pm0.12$ & $-4.8\pm3.6$ & \nodata & Y & G\\ 
 & F & $4$ & $15.8$ & $56$ & $-0.02\pm0.02$ & $-0.6\pm0.6$ & \nodata & Y & E\\
\enddata
\tablecomments{Columns are as follows: (1) IAU Name; $^\star$Source is a member of the representative sample;
$^\diamond$Motions for 0316$+$413 which are taken from unpublished data do not appear in Figure~1.  The motions for 0238$-$084
are from \cite{VRK03}.
(2) Component Identifier;  $^\dag$Component is the brightest.   
(3) Number of epochs with measured position; 
(4) Mean radial position, relative to the core; (5) Mean structural position
angle, relative to the core; (6) Angular radial speed and $1\sigma$ 
uncertainty; 
(7) Radial speed in units of the speed of light and $1\sigma$ uncertainty, 
computed assuming $H_0 = 70$~km\,s$^{-1}$\,Mpc$^{-1}$,
$\Omega_\mathrm{m}=0.3$, and $\Omega_\Lambda=0.7$;
(8) Extrapolated epoch of origin and uncertainty; 
(9) Does the component satisfy criterion ``b'' in Section 4.2, i.e., is it a well defined feature?; (10) Overall Quality Rating.}
\end{deluxetable*}

\clearpage

\begin{deluxetable*}{cccrrrrrr}
\tablecolumns{9}
\tablewidth{0pc}
\tabletypesize{\scriptsize}
\tablecaption{Vector proper motion results.\label{non-radial}}
\tablehead{
\colhead{Object} & \colhead{Comp} & \colhead{N} 
& \colhead{$<R>$} & \colhead{$<\vartheta>$} & \colhead{$\mu$} 
& \colhead{$\phi$} & \colhead{$\beta_{app}$} & \colhead{$|<\vartheta>-\phi|$}\\  
&&& \colhead{(mas)} & \colhead{(deg)}& \colhead{(mas/yr)} & \colhead{(deg)} & & \colhead{(deg)}\\
\colhead{(1)} & \colhead{(2)} & \colhead{(3)} & \colhead{(4)} &  
\colhead{(5)} & \colhead{(6)} & \colhead{(7)} & \colhead{(8)} & \colhead{(9)}   
}
\startdata 
$0035+413$
 & B & $5$ & $1.5\pm0.0$ & $103.7\pm0.8$ & $0.10\pm0.02$ & $121.8\pm8.2$ & $6.6\pm1.2$ & $18.1\pm8.2$\\ 
$0055+300$
 & B & $5$ & $0.9\pm0.0$ & $-54.2\pm1.6$ & $0.14\pm0.02$ & $-52.3\pm7.3$ & $0.2\pm0.0$ & $1.9\pm7.5$\\ 
$0106+013$
 & C & $5$ & $2.1\pm0.0$ & $-112.7\pm1.2$ & $0.27\pm0.03$ & $-119.6\pm7.4$ & $22.4\pm2.1$ & $6.9\pm7.5$\\ 
$0149+218$
 & C & $6$ & $5.0\pm0.1$ & $-21.9\pm1.1$ & $0.29\pm0.04$ & $-26.3\pm9.9$ & $18.7\pm2.3$ & $4.5\pm9.9$\\ 
${\bf 0234+285}$
 & {\bf B} & ${\bf 6}$ & ${\bf 3.5\pm0.1}$ & ${\bf -12.6\pm0.6}$ & ${\bf 0.25\pm0.05}$ & ${\bf 11.6\pm4.2}$ & ${\bf 14.7\pm2.8}$ & ${\bf 24.2\pm4.3}$\\ 
$0333+321$
 & B & $9$ & $2.7\pm0.0$ & $123.3\pm0.2$ & $0.18\pm0.01$ & $120.3\pm1.8$ & $11.0\pm0.4$ & $3.0\pm1.8$\\ 
 & C & $9$ & $4.9\pm0.0$ & $126.0\pm0.5$ & $0.20\pm0.02$ & $135.0\pm6.8$ & $12.2\pm1.4$ & $9.0\pm6.8$\\ 
$0336-019$
 & B & $5$ & $1.4\pm0.1$ & $66.9\pm1.8$ & $0.22\pm0.04$ & $67.1\pm7.0$ & $10.1\pm1.7$ & $0.2\pm7.2$\\
\enddata
\tablecomments{
Columns are as follows: (1) IAU Name (B1950); (2) Component
Identifier; (3) Number of epochs with measured position; (4) Mean
radial position, relative to the core; (5) Mean structural position
angle, relative to the core; (6) Angular velocity; (7) Direction of
velocity; (8) Linear speed in units of the speed of light;  (9) Misalignment
angle. Sources with significant non radial motion are shown in bold print.
}
\end{deluxetable*}



\begin{thebibliography}{}

\bibitem[Alef et al.(1996)]{AWK96}
Alef, W., Wu, S.~Y., Preuss, E., Kellermann, K.~I., \& Qiu, Y.~H.\
1996, \aap, 308, 376 

\bibitem[Aller et al.(1992)Aller, Aller, \& Hughes(1992)]{AAH92}
Aller, M.~F., Aller, H.~D., \& Hughes, P.~A.\ 1992, \apj, 399, 16

\bibitem[Aller et al.(2003)]{AAH03}
Aller, M.\ F., Aller, H.\ D., \& Hughes, P.~A.\ 2003, in ASP
Conf.\ Ser.\ 300, Radio Astronomy at the Fringe, ed.\ J.~A.\ Zensus,
M.~H.\ Cohen, \& E.\ Ros (San Francisco: ASP) 159


\bibitem[Barthel(1989)]{B89}
Barthel, P.~D.\ 1989, \apj, 336, 606 

\bibitem[Beasley et al.(2002)]{BGP02}
Beasley, A.~J., Gordon, D., Peck, A.~B., Petrov, L., MacMillan, D.~S.,
Fomalont, E.~B., \& Ma, C.\ 2002, \apjs, 141, 13 


\bibitem[Bennett et al.(2003)]{Ben03}
Bennett, C.~L., et al.\ 2003, \apjs, 148, 97


\bibitem[Blandford \& K\"onigl(1979)]{BK79}
Blandford, R.~D., \& K\"onigl, A.\ 1979, \apj, 232, 34 

\bibitem[Blandford \& Rees(1974)]{BR74}
Blandford, R.~D., \& Rees, M.~J.\ 1974, \mnras, 169, 395

\bibitem[Britzen et al.(1999)]{BVT99}
Britzen, S., Vermeulen, R.~C., Taylor, G.~B., Pearson, T.~J., Readhead,
A.~C.~S., Wilkinson, P.~N., \& Browne, I.~W.~A.\ 1999, in ASP Conf.\ Ser.\
159, BL~Lac Phenomenon, ed.\ L.~O.\ Takalo, A.\ Sillanp\"a\"a (San
Francisco: ASP), 431 

\bibitem[Britzen et al.(2001)]{BVT01}
Britzen, S., Vermeulen, R.~C., Taylor, G.~B., Campbell, R.~M., Browne,
I.~W.~A., Wilkinson, P.~N., Pearson, T.~J., \& Readhead, A.~C.~S.\ 2001,
in IAU Symp.\ 205, Galaxies and Their Constituents at the Highest
Angular Resolutions, ed.\ R.~T. Schilizzi, S.\ Vogel, F.\ Paresce, \& M.\
Elvis (San Francisco: ASP), 106 



\bibitem[Burbidge(2004)]{GRB04}
Burbidge, G.~R.\ 2004, \apj, in press


\bibitem[Carangelo et al.(2003)]{CFK03} Carangelo, N., 
Falomo, R., Kotilainen, J., Treves, A., \& Ulrich, M.-H.\ 2003, \aap, 412, 
651 


\bibitem[Cohen(1990)]{C90}
Cohen, M.~H.\ 1990, in Parsec-Scale Radio Jets, ed.\ J.~A. Zensus, \&
T.~J.\ Pearson (Cambridge: Cambridge University Press), 317.

\bibitem[Cohen et al.(1971)]{CCP71}
Cohen, M.~H., Cannon, W., Purcell, G.~H., Shaffer, D.~B., Broderick,
J.~J., Kellermann, K.~I., \& Jauncey, D.~L.\ 1971, \apj, 170, 207

\bibitem[Cohen et al.(1977)]{CLM77}
Cohen, M.~H., et al.\ 1977, \nat, 268, 405

\bibitem[Cohen et al.(1988)]{CBP88} 
Cohen, M.~H., Barthel, P.~D., Pearson, T.~J., \& Zensus, J.~A.\ 1988, \apj, 
329, 1 

\bibitem[Cohen et al.(2003)]{CRH03}
Cohen, M.~H., et al.\ 2003, in ASP Conf.\ Ser.\ 300, Radio Astronomy
at the Fringe, ed.\ J.~A.\ Zensus, M.~H.\ Cohen, \& E.\ Ros (San
Francisco: ASP), 27

\bibitem[Dallacasa et al.(2000)]{DSC00}
Dallacasa, D., Stanghellini, C., Centonza, M., \& Fanti, R.\ 2000,
\aap, 363, 887 

\bibitem[Dennett-Thorpe, Barthel, \& van Bemmel(2000)]{DBB00} 
Dennett-Thorpe, J., Barthel, P.~D., \& van Bemmel, I.~M.\ 2000, \aap, 364, 
501 

\bibitem[Dent(1965)]{D65}
Dent, W.~A.\ 1965, Science, 148, 1458

\bibitem[Dent(1972)]{D72}
Dent, W.~A.\ 1972, Science, 175, 1105

\bibitem[Dermer(1995)]{Dermer95}
Dermer, C.~D.\ 1995, \apjl, 446, L63 

\bibitem[Dermer \& Schlickeiser(1994)]{DS94}
Dermer, C.~D., \& Schlickeiser, R.\ 1994, \apjs, 90, 945

\bibitem[de Vries, Barthel, \& O'Dea(1997)]{DBO97}
de Vries, W.~H., Barthel, P.~D., \& O'Dea, C.~P.\ 1997, \aap, 321, 105



\bibitem[Ekers \& Liang(1990)]{EL90}
Ekers, R., \& Liang, H.\ 1990, in Parsec Scale Radio Jets, ed. J.~A.\
Zensus, \& T.~J. Pearson (Cambridge: Cambridge University Press), 333

\bibitem[Ghisellini et al.(1993)]{GPC93}
Ghisellini, G., Padovani, P., Celotti, A., \& Maraschi, L.\ 1993, \apj,
407, 65

\bibitem[Guijosa \& Daly(1996)]{GD96}
Guijosa, A., \& Daly, R.~A.\ 1996, \apj, 461, 600

\bibitem[Hartman et al.(1999)]{HBB99}
Hartman, R.~C., et al.\ 1999, \apjs, 123, 79 

\bibitem[Hirabayashi et al.(1998)]{H98}
Hirabayashi, H., et al.\ 1998, Science, 281, 1825
 

\bibitem [Homan et al.(2003)]{HLK03}
Homan, D.~C., Lister, M.~L., Kellermann, K.~I., Cohen, M.~C., Ros, E.,
Zensus, J.~A., Kadler, M., \& Vermeulen, R.~C. 2003, \apjl, 58, L9

\bibitem[Homan et al.(2001)]{HOW01}
Homan, D.~C., Ojha, R., Wardle, J.~F.~C., Roberts, D.~H., Aller, M.~F.,
Aller, H.~D., \& Hughes, P.~A.\ 2001, \apj, 549, 840



\bibitem[Hoyle(1966)]{HBS66}
Hoyle, F., Burbidge, G.~R., \& Sargent, W.\ 1966, \nat, 209, 751


\bibitem[Hughes et al.(1992)Hughes, Aller, \& Aller(1992)]{HAA92}
Hughes, P.~A., Aller, H.~D., \& Aller, M.~F.\ 1992, \apj, 396, 469

\bibitem[Istomin \& Pariev(1996)]{IP96}
Istomin, Y.~N., \& Pariev, V.~I.\ 1996, \mnras, 281, 1 


\bibitem[Jorstad et al.(2001a)]{JMMW01}
Jorstad, S.~G., Marscher, A.~P., Mattox, J.~R., Wehrle, A.~E., Bloom,
S.~D., \& Yurchenko, A.~V.\ 2001a, \apjs, 134,181

\bibitem[Jorstad et al.(2001b)]{JMMA01}
Jorstad, S.~G., Marscher, A.~P., Mattox, J.~R., Aller, M.~F., Aller,
H.~D., Wehrle, A.~E., \& Bloom, S.~D.\ 2001b, \apj, 556, 738 


\bibitem[Kellermann(2002)]{K02}
Kellermann, K.~I.\ 2002, PASA, 19, 77

\bibitem[Kellermann \& Pauliny-Toth(1969)]{KP69}
Kellermann, K.~I., \& Pauliny-Toth, I.~I.~K.\ 1969, \apjl, 155, L71

\bibitem[Kellermann et al.(1998a)]{K98A}
Kellermann, K.~I., Vermeulen, R.~C., Zensus, J.~A., \& Cohen, M.~H.\
1998, \aj, 115, 1295
	

\bibitem[Kellermann et al.(2000)]{K00}
Kellermann, K.~I., Vermeulen, R.~C., Zensus, J.~A., Cohen, M.~H.\ 2000,
in Astrophysical Phenomena Revealed by Space VLBI, ed.\ H.\ Hirabayashi,
P.~G.\ Edwards, \& D.~W. Murphy (Sagamihara, Japan: Institute of Space \&
Astronautical Science), 159

\bibitem[Kellermann et al.(1999)]{KVZ99}
Kellermann, K.~I., et al.\ 1999, New Astron.\ Reviews, 43, 757

\bibitem[Kellermann et al.(2003)]{Kel03}
Kellermann, K.~I., et al.\ 2003, in ASP Conf.\ Ser.\ 299, High Energy
Blazar Astronomy, ed.\ L.~O.\ Takalo, \& E.\ Valtaoja (San Francisco: ASP)
117


\bibitem[Kovalev(2003)]{K03}
Kovalev, Y.~Y.\ 2003, in ASP Conf.\ Ser.\ 300, Radio Astronomy at the
Fringe, ed.\ J.~A.\ Zensus, M.~H.\ Cohen, \& E.\ Ros (San Francisco:
ASP), 65

\bibitem[Kovalev(2004)]{K04}
Kovalev, Y.~Y.\ 2004, Baltic Astronomy, in press

\bibitem[Kovalev et al.(2002)]{KKNB02}
Kovalev, Y.~Y., Kovalev, Yu.~A., Nizhelsky, N.~A., \& Bogdantsov,
A.~V.\ 2002, PASA, 19, 83

\bibitem[Kovalev et al.(1999)]{KNK99}
Kovalev, Y.~Y., Nizhelsky, N.~A., Kovalev, Y.~A., Berlin, A.~B.,
Zhekanis, G.~V., Mingaliev, M.~G., \& Bogdantsov, A.~V.\ 1999, \aaps,
139, 545

\bibitem[K\"uhr et al.(1981)]{KWP81}
K\"uhr, H., Witzel, A., Pauliny-Toth, I.~I.~K., \& Nauber, U.\ 1981,
\aaps, 45, 367 

\bibitem[Lawrence(1996)]{Law96}
Lawrence, C.~L.\ 1996, in IAU Symp.\ 173, Astrophysical Applications of
Gravitational Lensing, ed.\ C.~S.\ Kochanek, \& J.~N.\ Hewitt (Dordrecht:
Kluwer), 299

\bibitem[L{\"a}hteenm{\"a}ki \& Valtaoja(1999)]{LV99}
L{\"a}hteenm{\"a}ki, A., \& Valtaoja, E.\ 1999, \apj, 521, 493 

\bibitem[L{\"a}hteenm{\"a}ki, Valtaoja, \& Wiik(1999)]{LVW99}
L{\"a}hteenm{\" a}ki, A., Valtaoja, E., \& Wiik, K.\ 1999, \apj, 511,
112

\bibitem[Lind \& Blandford(1985)]{LB85} Lind, K.~R.~\& 
Blandford, R.~D.\ 1985, \apj, 295, 358 

\bibitem[Lister(1999)]{L99}
Lister, M.~L.\ 1999, Ph.D.\ thesis, Boston University. 


\bibitem[Lister(2001)]{L01}
Lister, M.~L.\ 2001, \apj, 562, 208

\bibitem[Lister(2003)]{L03}
Lister, M.~L.\ 2003, in ASP Conf.\ Ser.\ 300, Radio Astronomy at the
Fringe, ed.\ J.~A.\ Zensus, M.~H.\ Cohen, \& E.\ Ros (San Francisco: ASP),
71

\bibitem[Lister \& Marscher(1997)]{LM97}
Lister, M.~L., \& Marscher, A.~P.\ 1997, \apj, 476, 572

\bibitem[Lister \& Marscher(1999)]{LM99}
Lister, M.~L., \& Marscher, A.~P.\ 1999, Astroparticle Physics, 11, 65


\bibitem[Lister et al.(2002)]{LKP02}
Lister, M.~L., Kellermann, K.~I., \& Pauliny-Toth, I.~I.~K.\ 2002, in
Proceedings of the 6th European VLBI Network Symp., ed.\ E.\ Ros, R.~W.\
Porcas, A.~P.\ Lobanov, \& J.~A.\ Zensus (Bonn: MPIfR), 135
 
\bibitem[Lister et al.(2003)]{LKV03}
Lister, M.~L., Kellermann, K.~I., Vermeulen, R.~C., Cohen, M.~H.,
Zensus, J.~A., \& Ros, E.\ 2003, \apj, 584, 135 



\bibitem[Massaro et al.(2001)]{MMF01}
Massaro, E., Mantovani, F., Fanti, R., Nesci, R., Tosti, G., \&
Venturi, T.\ 2001, \aap, 374, 435 

\bibitem[Mattox et al.(2001)]{MHR01}
Mattox, J.~R., Hartman, R.~C., \& Reimer, O.\ 2001, \apjs, 135, 155

\bibitem[Mattox et al.(1997)]{MSM97}
Mattox, J.~R., Schachter, J., Molnar, L., Hartman, R.~C., \& Patnaik,
A.~R.\ 1997, \apj, 481, 95




\bibitem[O'Dea(1998)]{OD98}
O'Dea, C.~P.\ 1998, \pasp, 110, 493

\bibitem[Orr \& Browne(1982)]{OB82}
Orr, M.~J.~L., \& Browne, I.~W.~A.\ 1982, \mnras, 200, 1067

\bibitem[Owsianik \& Conway(1998)]{OC98} Owsianik, I.~\& 
Conway, J.~E.\ 1998, \aap, 337, 69 

\bibitem[Padovani \& Urry(1992)]{PU92}
Padovani, P., \& Urry, C.~M.\ 1992, \apj, 387, 449

\bibitem[Pauliny-Toth \& Kellermann(1966)]{PK66}
Pauliny-Toth, I.~I.~K., \& Kellermann, K.~I.\ 1966, \apj, 146, 634




\bibitem[Readhead(1994)]{R94}
Readhead, A.~C.~S.\ 1994, \apj, 426, 51

\bibitem[Rees(1966)]{R66}
Rees, M.~J.\ 1966, \nat, 211, 468

\bibitem[Rees(1967)]{R67}
Rees, M.~J.\ 1967, \mnras, 135, 345


\bibitem[Ros, Zensus, \& Lobanov(2000)]{RZL00}
Ros, E., Zensus, J.~A., \& Lobanov, A.~P.\ 2000, \aap, 354, 55

\bibitem[Ros et al.(2002)]{RPL02}
Ros, E., Kellermann, K.~I., Lister, M.~L., Zensus, J.~A., Cohen, M.~H.
Vermeulen, R.~C., Kadler, M., \& Homan, D.~C.\ 2002, in Proceedings of the 6th European VLBI Network
Symp., ed.\ E.\ Ros, R.~W.\ Porcas, A.~P.\ Lobanov, \& J.~A.\ Zensus
(Bonn: MPIfR), 105


\bibitem[Shaffer, Kellermann, \& Cornwell(1999)]{SKC99} 
Shaffer, D.~B., Kellermann, K.~I., \& Cornwell, T.~J.\ 1999, \apj, 515, 558 


\bibitem[Sholomitskii(1965)]{S65}
Sholomitskii, G.~B.\ 1965, \azh, 42, 673; \sovast, 9, 516

\bibitem[Singal \& Gopal-Krishna(1985)]{SG85} Singal, 
K.~A.~\& Gopal-Krishna 1985, \mnras, 215, 383 


\bibitem[Sowards-Emmerd, Romani, \& Michelson(2003)]{SRM03}
Sowards-Emmerd, D., Romani, R.~W., \& Michelson, P.~F.\ 2003, \apj,
590, 109

\bibitem[Stickel, Meisenheimer, \& K\"uhr(1994)]{SMK94}
Stickel, M., Meisenheimer, K., \& K\"uhr, H.\ 1994, \aaps, 105, 211

\bibitem[Stickel et al.(1996)]{SRK96} 
Stickel, M., Rieke, G.~H., Kuehr, H., \& Rieke, M.~J.\ 1996, \apj,
468, 556 




\bibitem[Taylor et al.(1996)]{TVR96}
Taylor, G.~B., Vermeulen, R.~C., Readhead, A.~C.~S., Pearson, T.~J.,
Henstock, D.~R., \& Wilkinson, P.~N.\ 1996, \apjs, 107, 37 





\bibitem[Urry \& Padovani(1995)]{UP95}
Urry, C.~M., \& Padovani, P.\ 1995, \pasp, 107, 803 

\bibitem[Vermeulen(1995)]{V95}
Vermeulen, R.~C.\ 1995, PNAS, 92, 11385

\bibitem[Vermeulen \& Cohen(1994)]{VC94}
Vermeulen, R.~C., \& Cohen, M.~H.\ 1994, \apj, 430, 467 

\bibitem[Vermeulen et al.(2003a)]{VRK03}
Vermeulen, R.~C., Ros, E., Kellermann, K.~I., Cohen, M.~H., Zensus,
J.~A., \& van Langevelde, H.~J.\ 2003a, \aap, 401, 113

\bibitem[Vermeulen et al.(2003b)]{VBT03}
Vermeulen, R.~C., Britzen, S., Taylor, G.~B., Pearson, T.~J., Readhead,
A.~C.~S., Wilkinson, P.~N., \& Browne, I.~W.~A. 2003b, in ASP Conf.\
Ser.\ 300, Radio Astronomy at the Fringe, ed.\ J.~A.\ Zensus, M.~H.\
Cohen, \& E.\ Ros (San Francisco: ASP), 43

\bibitem[V{\' e}ron-Cetty \& V{\' e}ron(2001)]{VCV01}
V{\'e}ron-Cetty, M.-P., \& V{\' e}ron, P.\ 2001, \aap, 374, 92 

\bibitem[von Montigny et al.(1995)]{VM95}
von Montigny, C., et al.\ 1995, \apj, 440, 525


\bibitem[Walker et al.(2000)]{WDR00}
Walker, R.~C., Dhawan, V., Romney, J.~D., Kellermann, K.~I., \&
Vermeulen, R.~C.\ 2000, \apj, 530, 233



\bibitem[Wall \& Jackson(1997)]{WJ97}
Wall, J.~V., \& Jackson, C.~A.\ 1997, \mnras, 290, L17

\bibitem[Whitney et al.(1971)]{W71}
Whitney, A.~R., et al.\ 1971, Science, 173, 225



\bibitem[Woltjer(1966)]{W66}
Woltjer, L.\ 1966, \apj, 146, 597

\bibitem[Zensus, Cohen, \& Unwin(1995)]{ZCU95}
Zensus, J.~A., Cohen, M.~H., \& Unwin, S.~C.\ 1995, \apj, 443, 35 

\bibitem[Zensus et al.(2002)]{Z02}
Zensus, A., Ros, E., Kellermann, K.~I., Cohen, M.~H., Vermeulen, R.~C.\
2003a, \aj, 124, 662

\bibitem[Zensus et al.(2003)]{Z03}
Zensus, A., Ros, E., Kadler, M., Kellermann, K.~I., Lister, M.~L.,
Homan, D.~C., Cohen, M.~H., Vermeulen, R.~C.\ 2003b, in ASP Conf.\ Ser.\
300, Radio Astronomy at the Fringe, ed.\ J.~A.\ Zensus, M.~H.\ Cohen,
\& E.\ Ros (San Francisco: ASP), 27

\end{thebibliography}
\end{document}